\documentclass[]{emulateapj}
\usepackage{graphicx}
\usepackage{natbib}
\usepackage{amsmath}

\begin{document}

\title{The Diffuse Galactic Far-Ultraviolet Sky}
\author{Erika~T.~Hamden, David~Schiminovich}
\affil{Department of Astronomy, Columbia University,  550 W 120th St, New York, NY 10025, USA}
\email{hamden@astro.columbia.edu}
\and
\author{Mark~Seibert}
\affil{Carnegie Institution of Science, 813 Santa Barbara Street, Pasadena, California, 91101,USA}

\begin{abstract}

We present an all sky map of the diffuse Galactic far ultraviolet (1344-1786 \AA) background using GALEX data, covering 65\% of the sky with 11.79 arcmin$^2$ pixels.  We investigate the dependence of the background on Galactic coordinates, finding that a standard cosecant model of intensity is not a valid fit.  Furthermore, we compare our map to Galactic all sky maps of 100 $\mu$m emission, N$_{\rm HI}$ column, and H$\alpha$ intensity.  We measure a consistent low level FUV intensity at zero-points for other Galactic quantities, indicating a 300 CU non-scattered isotropic component to the diffuse FUV.  There is also a linear relationship between FUV and 100 $\mu$m emission below 100 $\mu$m values of 8 MJy/sr.  We find a similar linear relationship between FUV and N$_{\rm HI}$ below 10$^{21}$ cm$^{-2}$.  The relationship between FUV and H$\alpha$ intensity has no such constant cutoff.  For all Galactic quantities, the slope of the linear portion of the relationship decreases with Galactic latitude.  A modified cosecant model, taking into account dust scattering asymmetry and albedo, is able to accurately fit the diffuse FUV at latitudes above 20$^\circ$.  The best fit model indicates an albedo, $a$, of 0.62 $\pm$ 0.04 and a scattering asymmetry function, $g$, of 0.78 $\pm$ 0.05.  Deviations from the model fit may indicate regions of excess FUV emission from fluorescence or shock fronts, while low latitude regions with depressed FUV emission are likely the result of self-shielding dusty clouds.  

\end{abstract}


\section{Introduction}

Dust scattering of starlight by UV bright stars in our own Milky Way is now known to explain the majority of the diffuse far ultraviolet (FUV, $\sim$1300-1800 \AA) background.  However, interest in the diffuse FUV background began not with a focus on Galactic dust, but with a search for a cosmologically significant signal from the intergalactic medium (IGM) in order to quantify the total amount of mass and energy in the universe. \citet{1970Kurt} theorized that high energy photons emitted from dense hydrogen and helium in the IGM could be red shifted into the FUV, detectable as a diffuse isotropic continuum background.  A hot Galactic halo had also been proposed by \citet{1956Spitzer} as a source of diffuse FUV line emission. It is now well understood that these components do exist, although with lower surface brightness than initially theorized.

Among the early observations of the diffuse background were measurements by \citet{1976Morgan}, \citet{1980Paresce}, \citet{1984Jakobsen}, \citet{1989Fix}, \citet{1991Hurwitz}, \citet{1991Perault}, \citet{1999Murthy}.  While still incomplete in sky coverage, these observations hinted at a correlation between diffuse FUV intensity and Galactic neutral hydrogen column density.  This pointed to a Galactic source for the FUV, specifically scattering of UV star light by dust grains.  The Galactic origin of the diffuse FUV was more clearly determined with observations showing the correlation between diffuse FUV intensity and the infrared background at 100 $\mu$m as measured by the Infrared Astronomical Satellite (IRAS) \citep{1987Jakobsen,1991Perault,1996Sasseen}.  These results were further confirmed and expanded upon by recent missions with better sky coverage and angular resolution.  \citet{2001Schiminovich} observed one quarter of the sky in FUV with the Narrowband ultraviolet imaging experiment for wide-field surveys (NUVIEWS) rocket, comparing it to N$_{\rm HI}$ and 100 $\mu$m column, and finding a linear relationship at high latitudes.  Spectroscopic UV observations of parts of the diffuse sky have also been made by the Far-ultraviolet IMaging Spectrograph (FIMS/SPEAR) \citep{2006Edelstein} and the Far Ultraviolet Spectroscopic Explorer (FUSE) \citep{2000Moos}.  \citet{2011Seonb} use FIMS/SPEAR data and find correlations between FUV and 100 $\mu$m, N$_{\rm HI}$, and H$\alpha$.  \citet{2010Murthy} use low resolution Galaxy Evolution Explorer (GALEX) all sky data, with bright objects removed, and also find a strong correlation between FUV and 100 $\mu$m emission. 

While the correlation between diffuse FUV and dust column is now broadly accepted, deviations from this correlation are significant \citep{2011Seonb,2010Murthy,2001Schiminovich}. \citet{2004Murthy} found a weak correlation between FUV intensity and 100 $\mu$m using FUSE data, potentially due to differences in the local radiation field at low latitudes. On physical scales corresponding to molecular clouds, there can be significant deviations in the relationship between FUV and dust.  Observations in Aquila with FIMS/SPEAR \citep{2012Park} found FUV intensity correlates well with dust column for low extinction sightlines, while there is no correlation in regions with higher dust column.  Similarly, in the Draco Cloud, \citet{2010Sujatha} found substantial variations in the relationship between diffuse FUV intensity and 100 $\mu$m intensity using GALEX data.  The UV/IR ratio varied by a factor of 10 across the cloud.  Such divergent behaviors indicate that dust column is not the sole predictor of diffuse FUV intensity.  \citet{2013Seon} and \citet{2011Seonb} explain large variations in the UV/IR ratio as a result of a turbulent interstellar medium (ISM) which is represented as a lognormal function, where the standard deviation of a quantity increases with the mean value.

In the low density ISM, light from UV-bright stars (mostly near the plane of the Galaxy) is scattered off of dust grains, resulting in a low level diffuse FUV brightness which is correlated with dust content.  Above a certain threshold density, regions of the ISM may not reflect as much, as thicker clumps of dust attenuate FUV radiation.  \citet{2008Witt} and \citet{2011Seonb} find that this shielding begins at 100 $\mu$m $> 8$ MJy/sr, but at a range of FUV values.  Additionally, deviations from the FUV-dust correlation may indicate regions of especially high FUV radiation from nearby stars, a region of dust with unusual scattering properties, or even regions where molecular hydrogen is able to form and fluoresce.

Here we present a nearly all sky survey of the diffuse Galactic FUV background and compare the FUV intensity to 100 $\mu$m emission, N$_{\rm HI}$ observations, and H$\alpha$ intensity maps.  We employ a masking and mosaicing technique to remove FUV bright sources from all-sky survey images and create a composite map of the GALEX diffuse FUV sky.  This map provides unprecedented, wide and deep coverage compared to results from previous missions.  

We use this all sky data to investigate the precise nature of the relationship between FUV and tracers of cold Galactic dust and gas across the sky, focusing on how the relationship changes with both Galactic latitude and proximity to various Galactic plane associations.  The minimum FUV in these relations is also examined to determine if it reveals a significant isotropic extragalactic component, an un-modelled Galactic component, or another source.  The scatter in the relationships, both on large scales and within a single cloud, provide insight into the physical properties of the dust, including scattering asymmetry and albedo.  A clear picture of the FUV behavior and what drives it can also allow for the modeling and removal of the Galactic UV foreground.

In Section \ref{sec:data} we describe the data products used and any further analysis.  In Section \ref{subsec:mosaic} we describe the image mosaicing procedure and initial analysis of the GALEX data set in detail.  In Section \ref{sec:skydiscussion} we describe all sky trends and spatial distributions.  We discuss in particular the relationship between diffuse FUV and 100 $\mu$m emission (Section \ref{sec:slopes}) and H$\alpha$ intensity (Section \ref{sec:fuvhalpha}).  In Section \ref{sec:conclusion} we discuss the implications of our results.

\section{Data}
\label{sec:data}

GALEX, a 0.5 meter modified Ritchey-Chr$\acute{e}$tien telescope, operated for 10 years after its launch in 2003 \citep{2003M,2005M}.  GALEX observes in two UV channels (FUV (1344-1786 \AA) and NUV (1771-2831 \AA)) and has an angular resolution of 4.2 arcseconds (FWHM) in the FUV and 5.3 arcseconds (FWHM) in the NUV.  In this paper, we use data from the all sky survey (AIS), covering more than 25,000 square degrees on the sky with typical exposure time of 100 seconds, reaching a limiting point-source magnitude (m$_{AB}$) of 19.9 \citep[5 $\sigma$ AB;][]{2007Morrissey}. Each pointing center was chosen to minimize the gaps between adjacent fields.   While GALEX avoided bright stars in the Galactic plane and other regions, there is good coverage at higher latitudes.  We use images from the 6th GALEX data release (GR6), which contains a total of 34,551 individual AIS pointings.  GR6 contains nearly all AIS FUV data taken during the GALEX mission.  

Maps of the whole sky at 100 $\mu$m were taken from \citet{1998Schlegel}.  This map of the sky and the corresponding E(B-V) dust extinction maps were made by combining Cosmic Background Explorer/Diffuse Infrared Background Experiment (COBE/DIRBE) data with IRAS sky survey atlas (ISSA) maps in such a way as to accurately measure 100 $\mu$m emission (without a zero-point offset), which was then also used to derive a column density of dust.  This technique is able to estimate the dust at all but the lowest Galactic latitudes and densest clouds to 10\% precision.  N$_{\rm HI}$ data of the whole sky was taken from NASA's Legacy Archive for Microwave Background Data (LAMBDA) data service.  The all sky neutral hydrogen column density information is an interpolation of two maps, \citet{1997Hartmann} and \citet{1990Dickey}.  The \citet{1997Hartmann} map is a velocity integrated (-450 km/s $<$ V$_{lsr}$ $<$ +400 km/s) N$_{\rm HI}$ brightness temperature map sampled every 0.5$^\circ$ and converted to N$_{\rm HI}$.  The \citet{1990Dickey} map is a composite of several surveys averaged into 1 degree bins in Galactic coordinates with emission from -250 km/s $<$ V$_{lsr}$ $<$ +250 km/s.  H$\alpha$ data is taken from \citet{2003Finkbeiner} and has a 6' (FWHM) resolution.  It is a composite of the Virginia Tech Spectral line Survey (VTSS) in the northern hemisphere \citep{1998Dennison} and the Southern H$\alpha$ sky survey atlas (SHASSA) in the southern hemisphere \citep{1998Dennison}.  The Wisconson H$\alpha$ Mapper (WHAM) \citep{2002Reynolds} provides a stable zero point at a 1 degree scale.

\subsection{Image Mosaicing}
\label{subsec:mosaic}

In order to observe the diffuse background intensity, we create high resolution FUV images with known point and resolved sources removed.  To create these mosaics, we use the GALEX data products described in \citet{2007Morrissey} along with the Montage software package \citep{2003Berriman,2005Laity}.  

In our analysis we use four main maps to generate FUV background images: \textit{cnt} (counts per pixel), \textit{rrhr} (relative response or effective exposure time per pixel), \textit{sky} (estimated sky background), and \textit{mask} (detected objects).  The sky background file is created by the GALEX pipeline and is an estimate of the smoothed background after resolved and point sources are removed from each image \citep{2007Morrissey}. The \textit{mask} file provides the locations of pixels that contain UV-detected objects, which are removed for background estimation.  The flagged pixels in this pipeline mask file---also called a segmentation file by the Sextractor object-detection software used to perform photometry on GALEX images---only contains contiguous pixels from an object that are well detected above background, and may not include extended faint light (or optical ghosts, etc.) associated with an object.

Our mosaicing procedure involves several steps.  First we remove resolved sources from each file to be mosaiced, using the \textit{mask} and \textit{cnt} files.  The \textit{mask} file is smoothed using a boxcar of width 10 x 10 pixels to place an extra 15\arcsec\ border around the objects being masked.  This is done to more effectively block light from bright stars and galaxies, which can extend beyond the unsmoothed masked area.  Even with this extra border a fraction of the light from an object will remain unmasked.  Encircled energy curves for the GALEX FUV PSF indicate that 5\% of the light extends beyond 20 arcseconds radius \citep{2007Morrissey}, our typical minimum masked radius. Bright objects will usually have an even larger extent in the object mask, thereby reducing this fraction. 

\begin{figure}
\centering
\includegraphics[width=0.45\textwidth]{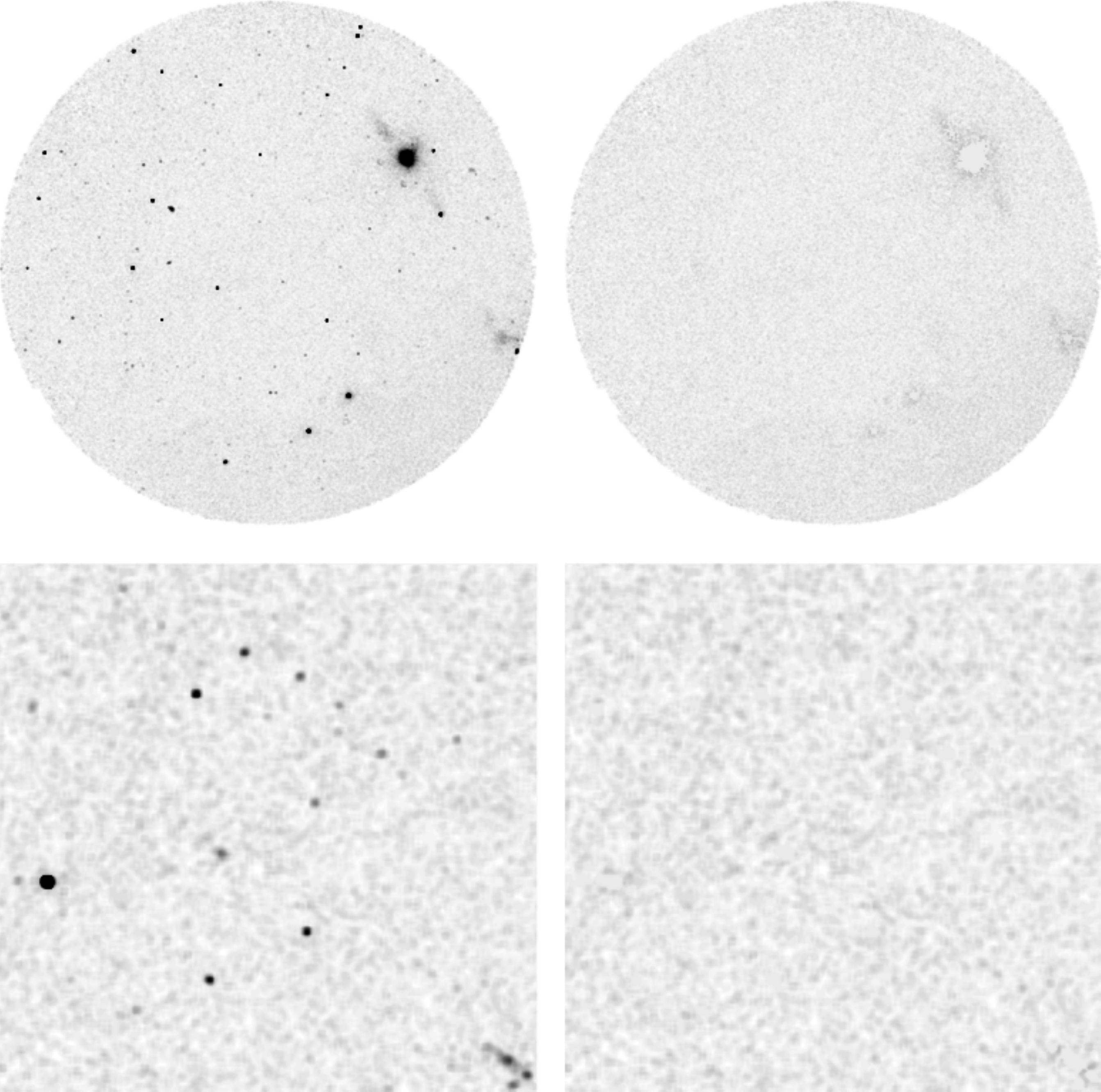}
\caption{Examples of GALEX image masking.  \textbf{Top Left:} Image of an unmasked GALEX field with a particularly bright A0 star (V=9.48, FUV=12.75, saturated) on the right of the field.  \textbf{Top Right:} Image of same field after masking.  Regions covered by the mask have been replaced by the estimated sky background from the \textit{sky} images \citep{2007Morrissey}.  Some reflections from bright stars remain after masking.  \textbf{Bottom Left:} Close up image of an unmasked GALEX field with dimmer point sources.  \textbf{Bottom Right:} Image of same field after masking, leaving behind a smooth background.}
\label{fig:mask}
\end{figure}

For display purposes, the flagged areas on the smoothed \textit{mask} are then excised from the \textit{cnt} files and are replaced with the corresponding section of the \textit{sky} file, which was generated by the pipeline as an estimation of the background in that region.  Given the AIS source density, this background replacement has only a small impact on the overall noise in the images.  The \textit{sky} file is in units of counts per second, so we multiply the \textit{sky} file by the corresponding regions of the \textit{rrhr} file to maintain the correct units of counts.  This can be described by:

\begin{equation}
\begin{aligned}
cnt_{masked,i,j} =& mask_{i,j}*sky_{i,j}*rrhr_{i,j} \\
        &+ (1-mask_{i,j})*cnt_{i,j} 
\end{aligned}
\end{equation}

\noindent where mask$_{i,j}=1$ for detected objects. Figure \ref{fig:mask} shows two different GALEX images before and after this procedure.  

The next step was to create a set of GALEX FUV mosaics centered on 12,288 equally spaced points covering the whole sky.  Each point was taken from a nested Hierarchical Equal Area isoLatitude Pixelization (HEALPix) ordering to evenly cover the sky \citep{2005Gorski}, with Nside=32.   Each mosaiced image contains all GALEX AIS fields within a 3 degree radius from the center of the pointing, with some overlap between neighboring images.  Using Montage, the \textit{rrhr} and masked \textit{cnt} files are reprojected so all images to be mosaiced lie in the same plane.  The overall size of the \textit{cnt} file is also trimmed to remove the edges.  Reprojected files are then mosaiced into large \textit{rrhr} and \textit{cnt} files.  The final step is to divide the mosaiced \textit{cnt} file by the mosaiced \textit{rrhr} file, creating a finished mosaic with units of counts per second.  

\begin{equation}
I_{cnts/sec} = \frac{cnt_{masked,mosaiced}}{rrhr_{mosaiced}}
\end{equation}

The image units of cnts/sec are then converted to photons cm$^{-2}$ sec$^{-1}$ sr$^{-1}$ \AA$^{-1}$, hereafter referred to as continuum units (CU).  The conversion from counts/sec to flux is 1.40 $\times$ 10$^{-15}$ for FUV, with appropriate conversions from ergs cm$^{-2}$ to photons sr$^{-1}$.  The unit converted mosaic is then compared to the all sky maps described above.

\begin{figure}
\centering
\includegraphics[width=0.45\textwidth]{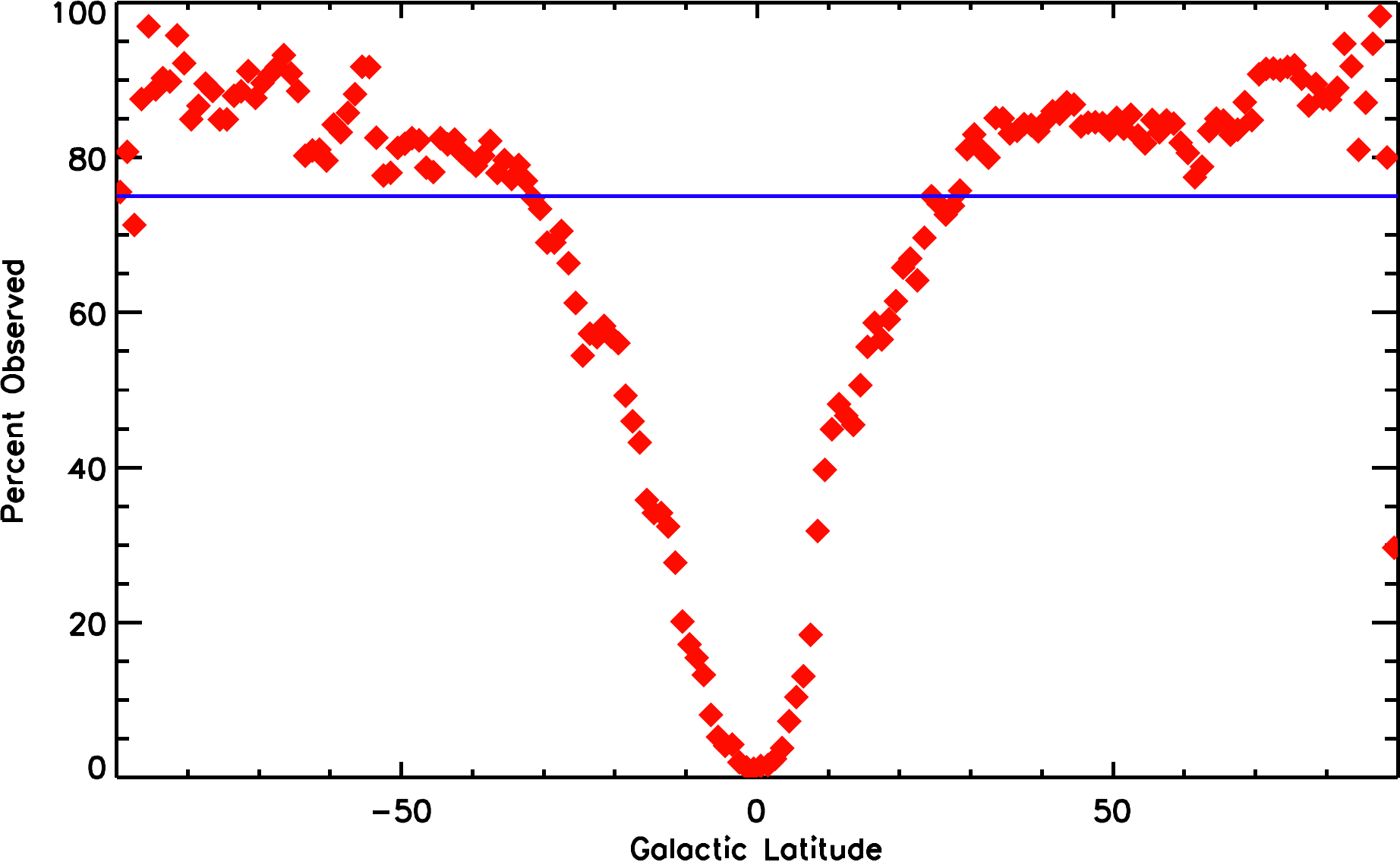}
\caption{Percentage of coverage of GALEX FUV AIS data for a given Galactic latitude in bins of 1$^\circ$.  Light blue line indicates 75\% coverage to guide the eye.  Regions near the Galactic plane have much lower coverage than higher latitudes.  The average coverage above $|$b$|$=25$^\circ$ is 81\%.}
\label{fig:percover}
\end{figure}

GALEX FUV data was not available for a fraction of these points, due to the avoidance of the Galactic plane and other bright objects.  The sensitivity of the GALEX detector limited the max count rate for FUV AIS observations to 5000 cnts/s \citep{2007Morrissey}.  Of the 12,288 points, 10,019 had GALEX AIS fields within the 3 degree radius, using a total of 28,938 individual GALEX fields.   Figure \ref{fig:percover} shows the percentage of the sky covered by our maps for a given Galactic latitude.  The lowest latitude regions ($|$b$| <$ 25$^\circ$) have coverage below 75\%.  There is a slight asymmetry between north and south hemispheres in this plot, due to the location of the Orion OB association below the Galactic plane.

Our final step was to create an all sky map, with each image described above sampled onto lower resolution pixels.  A total of 12,582,912 pixels cover the whole sky in a nested HEALPix ordering \citep{2005Gorski} with each pixel covering 11.79 arcmin$^2$ and Nside=1024.  

Assuming Poisson errors, with signal to noise equal to the square root of the signal, we find a typical AIS image (with 100 second exposure) will yield a signal to noise of $\sim$ 16 per HEALPix pixel.  Regions of the sky with more than one GALEX AIS pointing, overlapping edges of pointings, and averaging over larger areas will yield greater signal to noise values.  The all sky map is shown in Figure \ref{fig:fuvallsky}.  The map contains 65\% of the sky, as compared to 25\% from \citet{2001Schiminovich}, 75\% from \citet{2010Murthy}, and 80\% from \citet{2011Seonb}. 

\section{The Diffuse FUV Sky}
\label{sec:skydiscussion}

\begin{figure}
\centering
\includegraphics[width=0.45\textwidth]{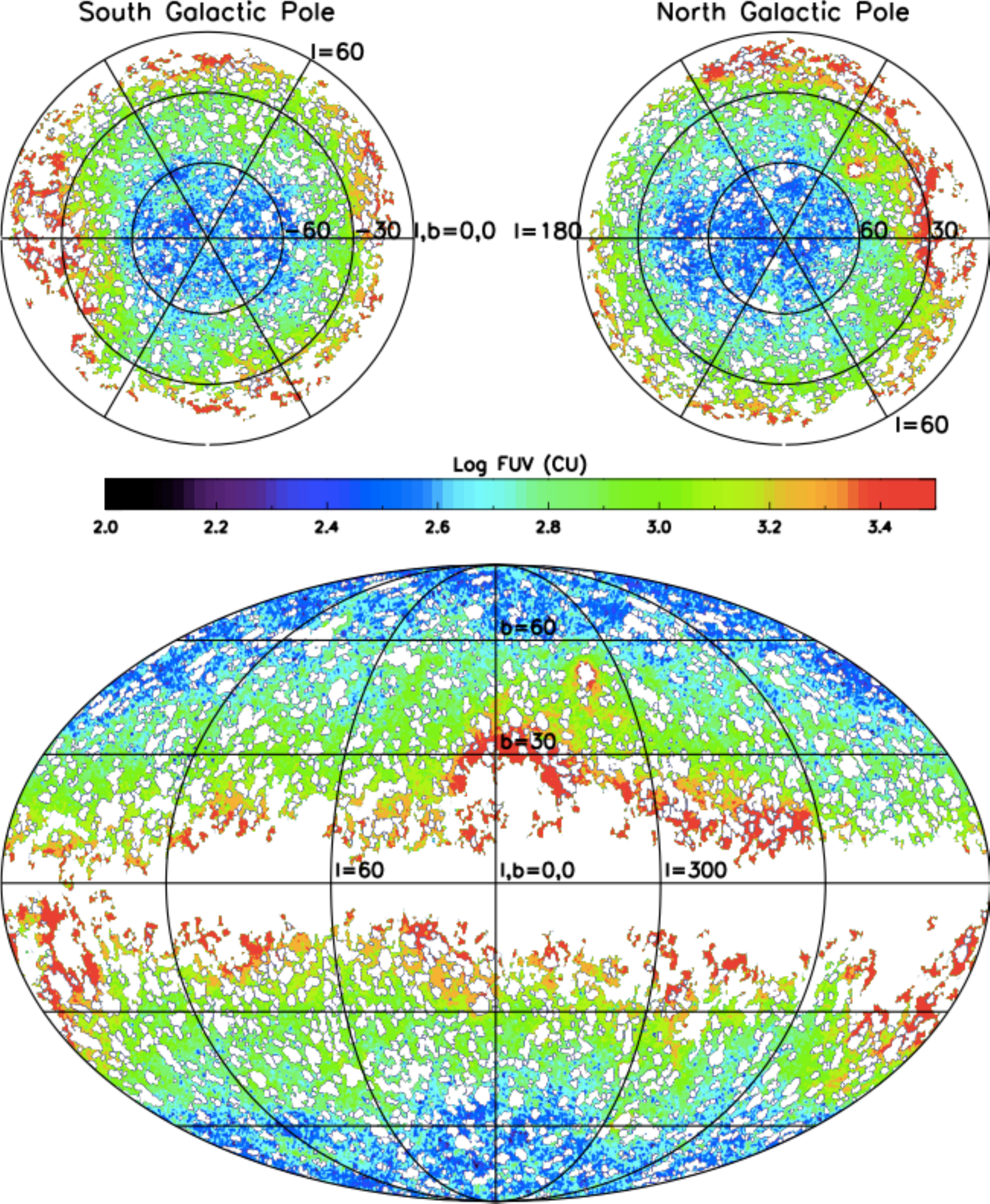}
\caption{Log of diffuse FUV intensity (CU) across the sky.  The lowest FUV intensity (a few hundred CU) is at the highest latitudes, while the highest FUV intensity (a few thousand CU) is found nearest to the Galactic plane.  The highest intensity observed is at the edges of known OB associations, near dense molecular clouds.  Overall intensity is best fit as a modified cosecant with latitude, as is discussed in Section \ref{sec:galactictrends}.}
\label{fig:fuvallsky}
\end{figure}

Here we present the GALEX diffuse FUV all sky map.  Figure \ref{fig:fuvallsky} shows the composite map in log scale.  At high latitudes, FUV intensity is low, reaching a lower limit of a few hundred CU.  At lower latitudes closer to the Galactic plane, the intensity increases to several thousand CU.  The growing intensity towards the plane follows a rough cosecant trend with latitude as discussed further below.   The highest intensities are found at the edges of known OB associations near dense molecular clouds:  Ophiuchus ($l$=355$^\circ$, $b$=18$^\circ$), the Tau-Per-Aur Complex ($l$=170$^\circ$, $b$=-15$^\circ$), and the Orion A \& B complex ($l$=200-220$^\circ$, $b$=-17$^\circ$) \citep{2001Dame}.

\begin{figure}
\centering
\includegraphics[width=0.45\textwidth]{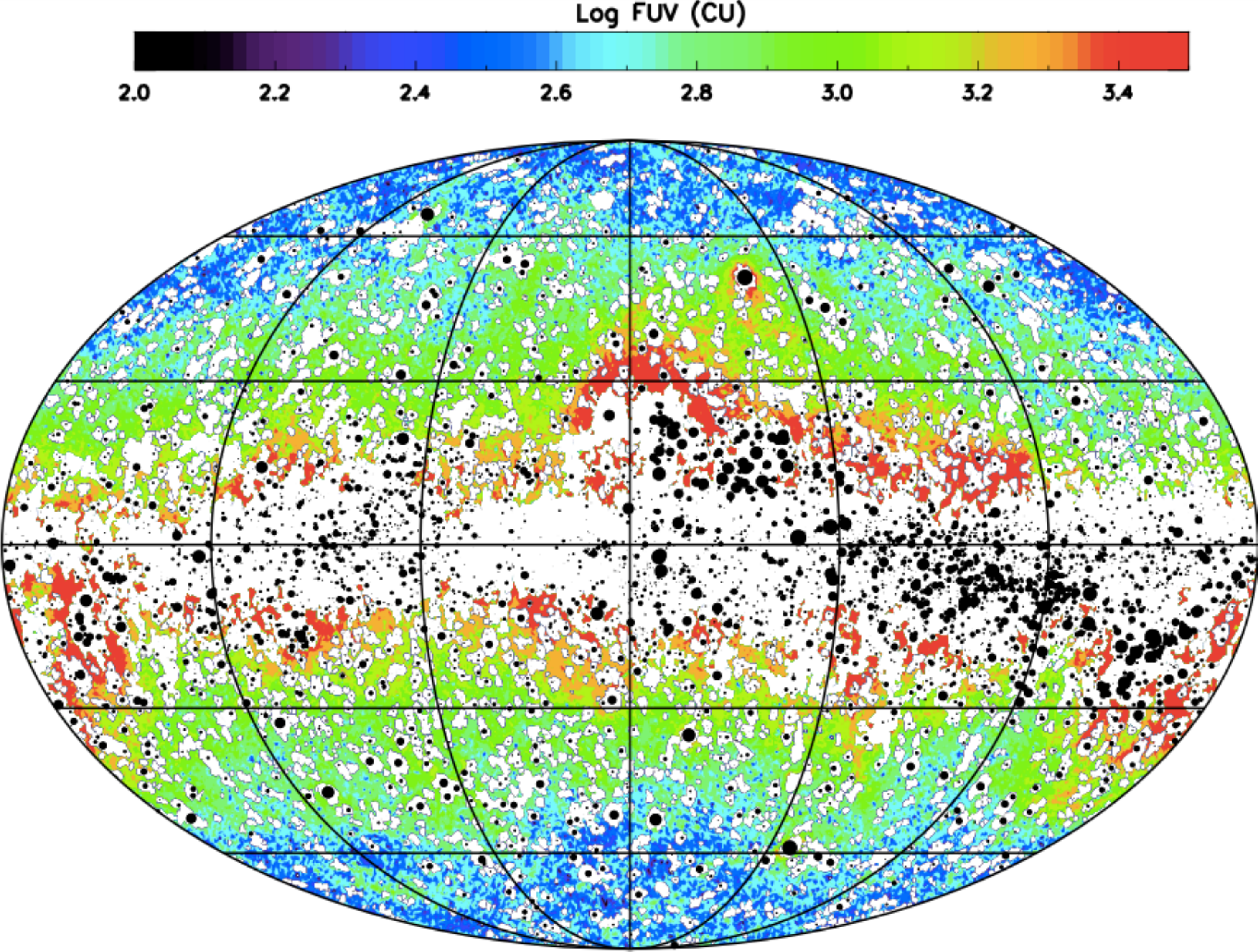}
\caption{Log of diffuse FUV intensity (CU) across the sky, with locations of TD-1 bright stars overplotted.  The diameter of the points is proportional to the log of the FUV flux.  The coordinates are as in Figure \ref{fig:fuvallsky}}
\label{fig:starlocations}
\end{figure}

\citet{2011Seonb} and \citet{2001Schiminovich} both note a ``significant depression'' in the FUV maps at latitudes above b$>$20$^\circ$ between $l$=20$^\circ$ to $l$=60$^\circ$.  Overall, we find the intensity at mid to high latitudes here is mostly consistent with intensity at similar latitude regions in other parts of the sky.  Regions of the sky with unusually high intensity ($>$ 5000 CU) can be linked to the OB associations mentioned above and are related to the uneven distribution of FUV bright stars.  Figure \ref{fig:starlocations} shows the positions of bright stars (Flux$_{1565 \text{\AA}} >$ 1 $\times$ 10$^{-12}$ erg cm$^{-2}$ s$^{-1}$ \AA$^{-1}$) from the TD-1 survey of nearby stars \citep{1978Thompson}.  Ophiuchus, the Orion complex, and other areas of high FUV emission have an excess of bright stars.

\subsection{Galactic Trends}
\label{sec:galactictrends}

\begin{figure}
\centering
\includegraphics[width=0.45\textwidth]{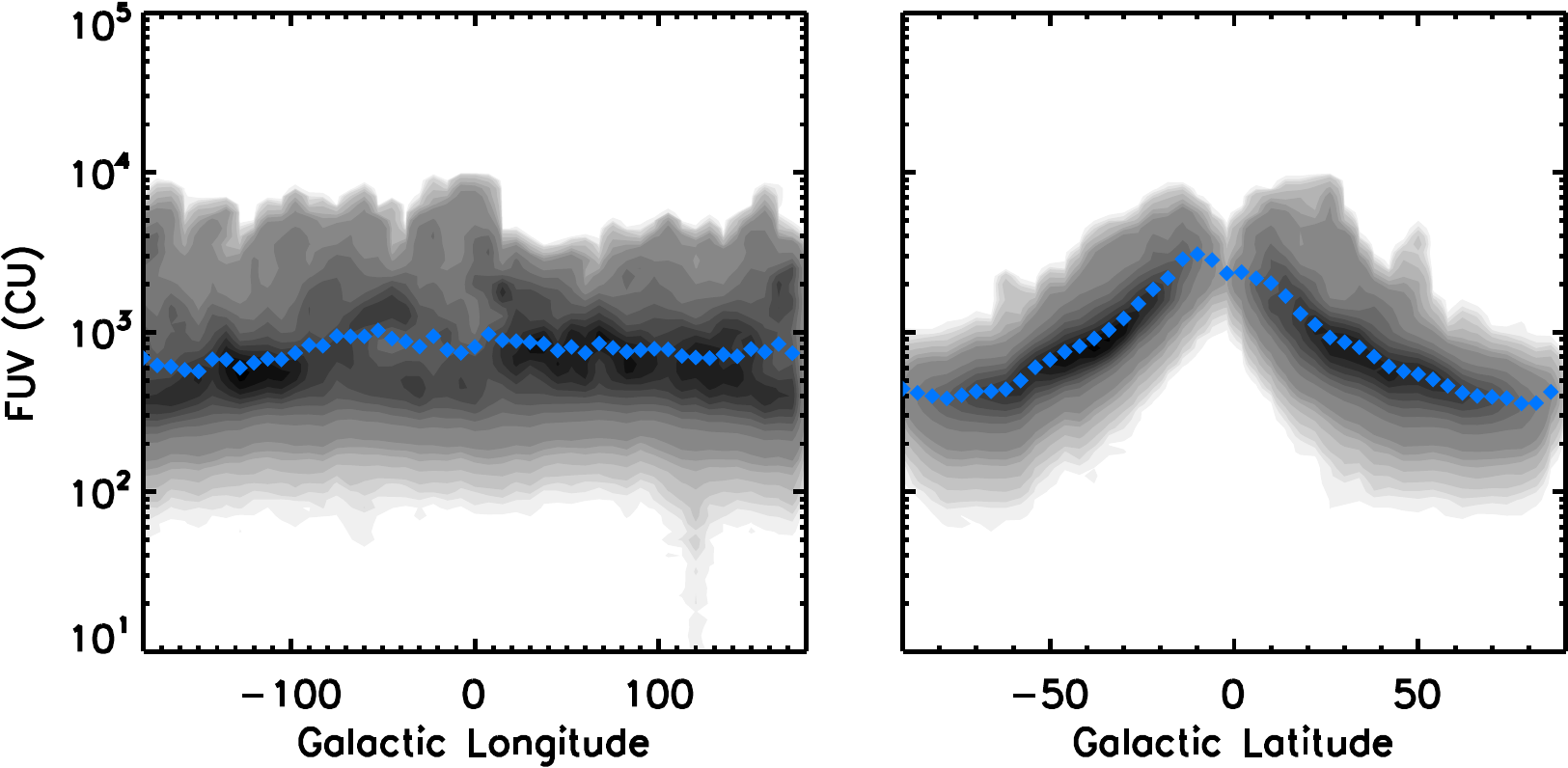}
\caption{\textbf{Left:} Plot of FUV vs. Galactic longitude.  Blue dots indicate median FUV intensity in 5$^\circ$ bins.  Median FUV intensity (800-1000 CU) is relatively constant across longitudes. \textbf{Right:} Plot of FUV vs. Galactic latitude.  Blue dots indicate median FUV intensity in 3$^\circ$ bins.  The lowest latitudes have fewer points since GALEX has not observed the entire Galactic plane. }\label{fig:fuvlat}
\end{figure}

FUV intensity vs. Galactic latitude and longitude are shown in Figure \ref{fig:fuvlat}.  The left panel shows diffuse FUV intensity vs. Galactic longitude.  Overall, we find this matches well with the diffuse FUV intensity from SPEAR/FIMS described in \citet{2011Seonb}.  The right panel shows FUV intensity vs. Galactic latitude.  FUV intensity increases with decreasing absolute value of the latitude, and appears to be relatively symmetric between northern and southern Galactic hemispheres.  The avoidance by GALEX of UV bright regions will bias the latitude-averaged intensity at the lowest latitudes ($|$b$|$ $<$ 20), evident when compared to \citet{2011Seonb} which reaches values of 10,000 CU in the plane. 

Certain Galactic quantities, including column densities and absorption, have been known to follow a cosecant shape with latitude, derived by \citet{1940Parenago} to model Galactic reddening.  This model has been expanded and refined, but the basic principle remains the same.  Following \citet{1980Milne} and \citet{1966Sturch}, Galactic extinction can be modeled as:

\begin{equation}
C = \int_{r=0}^d k_o \xi(z) dr = k_o \csc |b| \times \int_{0}^z \xi(z) dz
\label{eq:csc}
\end{equation}

\noindent where k$_o$ is reddening in the plane, $z$ is height above the plane, and $\xi(z)$ is a function that describes how reddening changes with $z$.  Using simple trigonometry, we replace a radial distance with $z = \sin|$b$| \times r$.  The resulting cosecant dependence is shown in the right hand side of Eq \ref{eq:csc}.  Several functions have been suggested for $\xi(z)$, including exponential with a scale height \citep{1980Milne}, although the exact form is not relevant here.  If $C$ traces the amount of obscuring dust and the scattered FUV is proportional to the dust column, then one can relate the two by a scale factor, k$_{scatter}$.

\begin{equation}
I_{FUV}= k_{scatter} \times C
\label{eq:final}
\end{equation}

With a latitude dependent extinction model, we can reasonably expect a latitude dependent FUV intensity.  This model has been fit by \citet{1991Perault} and others.

\begin{figure}
\centering
\includegraphics[width=0.45\textwidth]{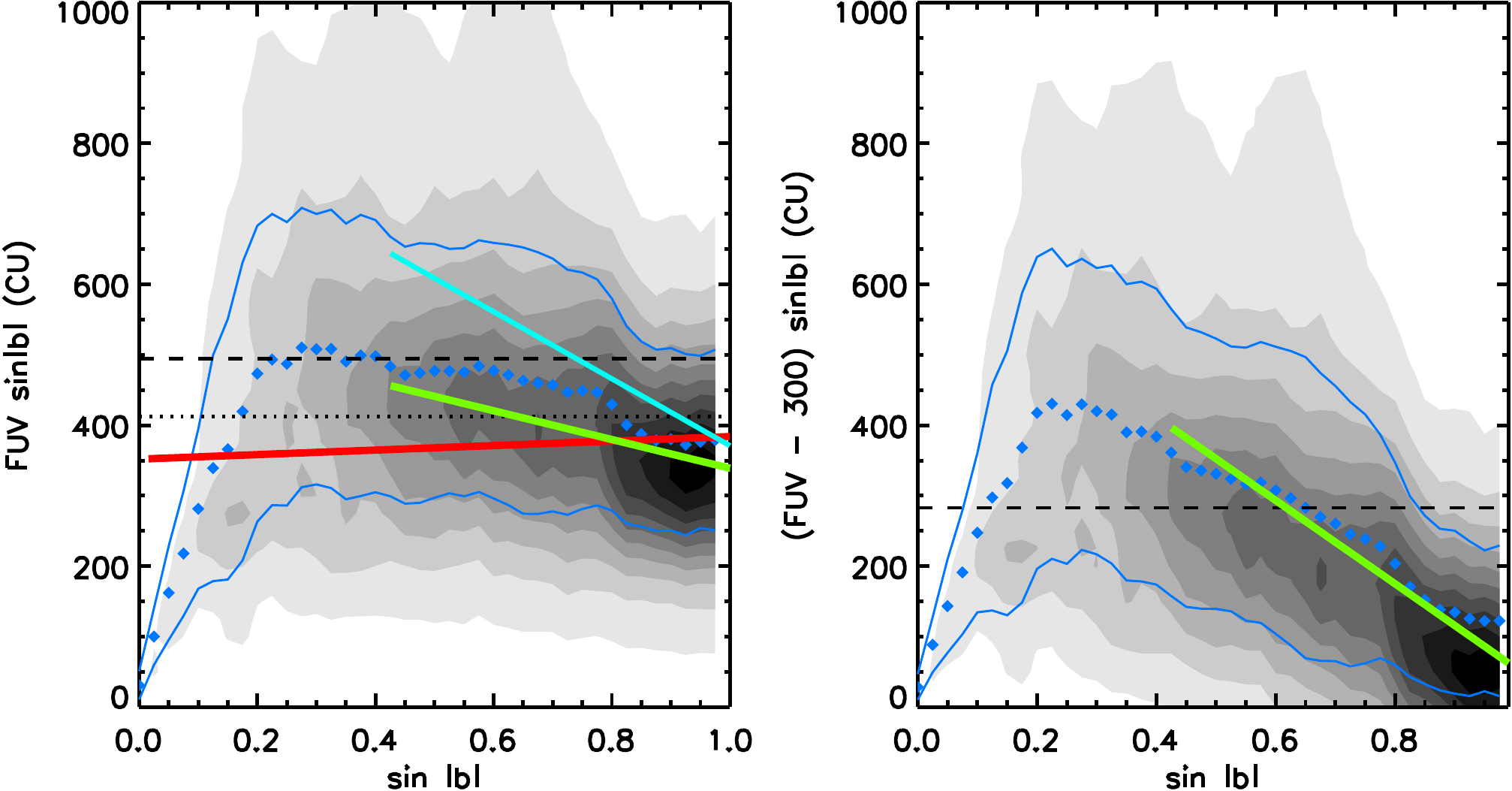}
\caption{Plot of FUV sin$|$b$|$ vs. sin$|$b$|$.  \textbf{Left:} 2-d histogram of FUV sin$|$b$|$ as a function of Galactic latitude.  The black dashed line is the single parameter fit to FUV sin$|$b$|$ at all points, 495 CU.  The black dotted line is the single parameter fit from \citet{2011Seonb}, 412 CU.  The green and red lines are best fits for a function of the form I=A/sin$|$b$|$ + B, with green fitting points above $|$b$|$=25$^\circ$ and red for all points.  The cyan line is the two parameter fit from \citet{2011Seonb}.  \textbf{Right} 2-d histogram of the same data as the top plots, but with a 300 CU offset removed.  The black dashed line is the median for all points, while the green line is a two parameter fit for $|$b$| >$ 25$^\circ$.}\label{fig:fuvsin}
\end{figure}

The cosecant dependence of FUV is shown in the left panel of Figure \ref{fig:fuvsin} which plots FUV sin$|$b$|$ vs. sin$|$b$|$.  Including all points, the median FUV sin$|$b$|$ is 451 CU, slightly lower than the value from \citet{2011Seonb}, of 525.4 CU.  This difference we ascribe to unobserved high intensity regions as discussed previously.  While the unobserved regions are concentrated in the plane, high intensity areas at all latitudes are not included in our data.  Thus an overall lower median for our all-sky data is to be expected.  The deviation from a constant value with sin$|$b$|$ at the lowest latitudes is probably due to the fact that the line of sight is no longer optically thin.  At higher latitudes, the behavior is roughly flat, consistent with a cosecant form, with a slight decrease above sin$|$b$|$=0.8.

Fitting a function of the form of Equation \ref{eq:final} corresponds to a horizontal line on this plot.  If we assume FUV=A/sin$|$b$|$, we find A=495.  This is the dashed line in the plot, compared to 412.3$\pm$10.3 (dotted line) from \citet{2011Seonb}.  This is not a good fit to our data at any latitude.  Adding an additional constant term and fitting a function of the form I = A/$\sin|b|$ + D, we find that the values for A and D vary significantly depending on which latitudes are included.  The red line shows the fit for all points with $|$b$| >$ 5 [A=352, D=32].  This fit fails to adequately capture the behavior at all latitudes.  Instead, fitting for points with $|$b$| >$ 25, shown as the green line in the left panel of Figure \ref{fig:fuvsin} [A=543, D=-204].  \citet{2011Seonb} find values of 847$\pm$96 and -457$\pm$100, for A and D respectively, shown in cyan.  These differences are again likely due to GALEX avoidance of bright objects.  At $\sin|b| = $1.0, the value of the FUV intensity will reduce to A+D, yielding a value for a minimum FUV intensity.  For both fits, this number is a few hundred CU.  

The two parameter fit, while relatively good for higher latitudes, raises the question about the physical interpretation of the value for D.  The value of A should indicate the scale factor from extinction or dust column to scattering, while a positive value for D could be interpreted as an isotropic component which is unrelated to the scattering traced by the cosecant fit.  If the isotropic component is already removed, or if D is negative, then there is no physical motivation for including this term.  Further discussion of and modification to the cosecant fit is found in Section \ref{sec:conclusion}.

We show the result of removing an offset in the right plot in Figure \ref{fig:fuvsin}.   We use 300 CU as a low level intensity that is not related to Galactic components.  Further discussion of this component is found in Sections \ref{sec:fuvvsall} and \ref{sec:conclusion}.  As noted above, a single parameter cosecant distribution will be horizontal.  Instead, the actual data decrease with increasing sin$|$b$|$, indicating that the cosecant model doesn't provide a reasonable fit when a physically motivated isotropic offset is removed.  The straight dashed line fit of the median (I$_{FUV}$-300) sin$|$b$|$ = 283 does not match the data at any latitude range.  A single parameter fit yields A=312, similar to the median.  Forcing a two parameter fit for $|$b$| >$ 25 yields values of [A=648, D=-592] which provides a better match, but has no straightforward physical interpretation.

\subsection{Relationship with other Galactic properties}
\label{sec:fuvvsall}

The diffuse FUV intensity at high latitudes is determined by the distributions and intensities of both FUV emission from bright stars throughout the disk and the dust which scatters that emission.  Here we investigate the relationship between FUV intensity and other Galactic quantities that trace dust and gas.  In this section all sky maps of 100 $\mu$m emission, H$\alpha$ intensity, and N$_{\rm HI}$ column density are each compared to the diffuse FUV data.   While there are overall correlations, we also explore how scatter may provide information about the distribution and properties of the dust and illuminating sources.  We note here that the resolution of the N$_{\rm HI}$ column density map is significantly lower than the other maps used.  We expect this will increase the scatter in our correlation between FUV and N$_{\rm HI}$ column, but will not change the overall result. 

\begin{figure}
\centering
\includegraphics[width=.45\textwidth]{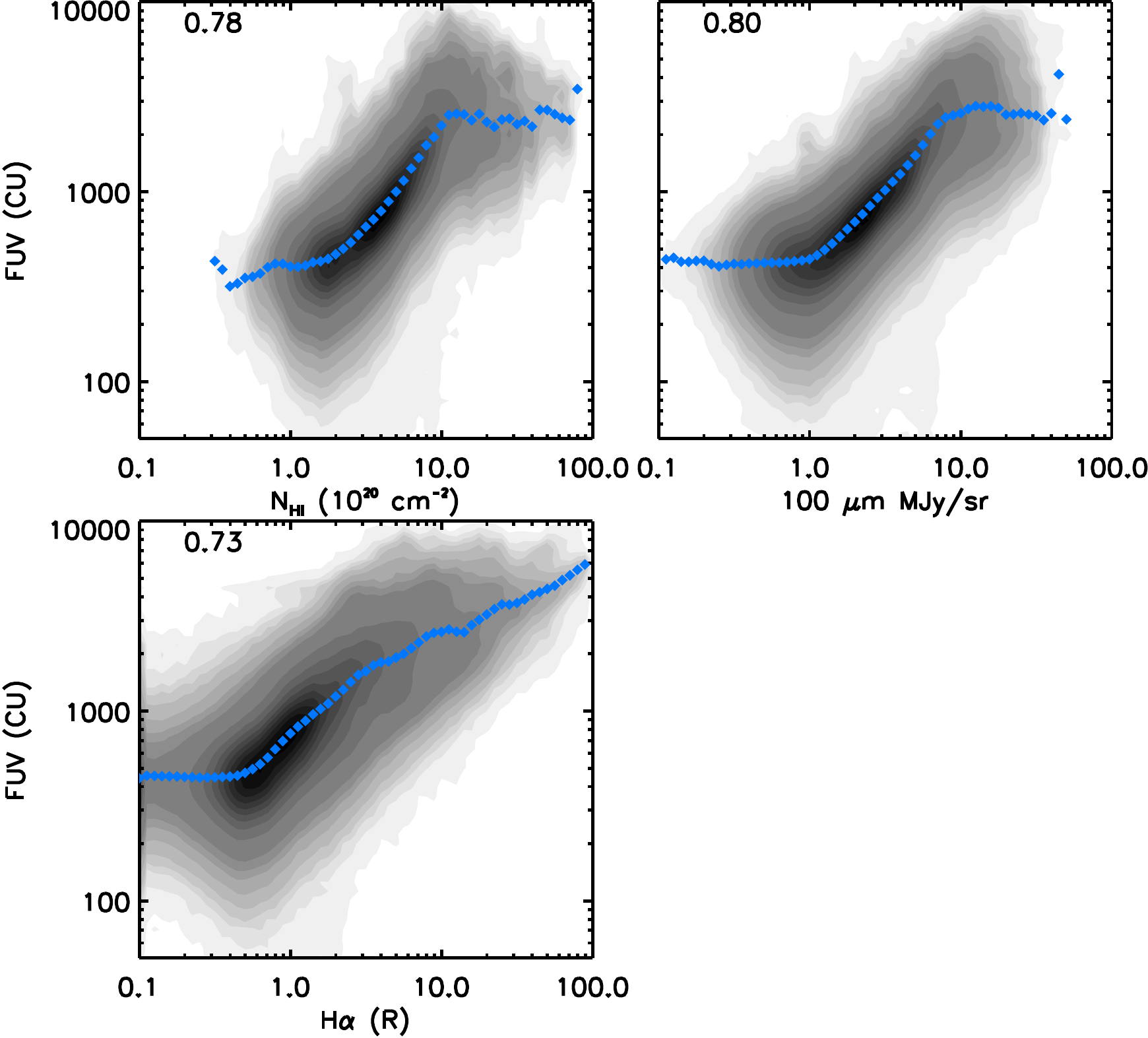}
\caption{2-D histogram of FUV intensity vs. Galactic quantities in log-log space.  Blue dots indicate the median value in abscissa bins of 0.05 dex.  The linear Pearson correlation coefficient is shown in the upper left of each panel.  \textbf{Top Left:} FUV intensity vs. N$_{\rm HI}$.  At low ($<$ 10$^{20}$ cm$^{-2}$) and high values ($>$ 10$^{21}$ cm$^{-2}$) of N$_{\rm HI}$, the FUV intensity levels off.  \textbf{Top Right:} FUV intensity vs. 100 $\mu$m emission.  As with N$_{\rm HI}$, the relationship is flat at low ($<$ 1 MJy/sr) and high ($>$ 8 MJy/sr) 100 $\mu$m intensity, with a linear relationship in between.  \textbf{Bottom Left:} FUV intensity vs. H$\alpha$ intensity.  Unlike the previous two plots, the FUV intensity does not level off at high values of H$\alpha$ intensity.}\label{fig:fuvvsglobal}
\end{figure}

Two dimensional histograms are plotted for FUV emission vs. each Galactic quantity in log-log space in Figure \ref{fig:fuvvsglobal}.  All three graphs show strong correlation between the FUV intensity and the other measured quantities (correlation values calculated using the linear Pearson method are shown).  As found by \citet{2011Seonb}, all three quantities are well correlated with FUV emission, but include a large amount of scatter.  The strongest correlation is between FUV emission and 100 $\mu$m intensity, with r=0.80.  The correlation between FUV emission and N$_{\rm HI}$ is also quite high, with r=0.78.  The weakest correlation is between FUV emission and H$\alpha$ intensity, with r=0.73.  

In all three plots of Figure \ref{fig:fuvvsglobal}, we find a low level minimum FUV.  The FUV intensity has a minimum at around a few hundred CU, flattening below 2 $\times$10$^{20}$ cm$^{-2}$ for N$_{\rm HI}$, 1 MJy/sr for 100 $\mu$, and 0.5 Rayleighs (R) for H$\alpha$.  The plots of N$_{\rm HI}$ and 100 $\mu$m both also have a significant flattening of FUV intensity at large values.  The FUV median remains constant above 10 $\times$10$^{20}$ cm$^{-2}$ for N$_{\rm HI}$ and 8 MJy/sr for 100 $\mu$.  The plot of FUV vs. H$\alpha$, however, continues to be linear at high values of both, also seen in \citet{2011Seona}.  Some flattening at large values of 100 $\mu$m and N$_{\rm HI}$ was also observed by \citet{2011Seonb} with SPEAR/FIMS data.  The GALEX avoidance of bright regions of the sky could make this more pronounced in our data.

\begin{figure}
\centering
\includegraphics[width=0.45\textwidth]{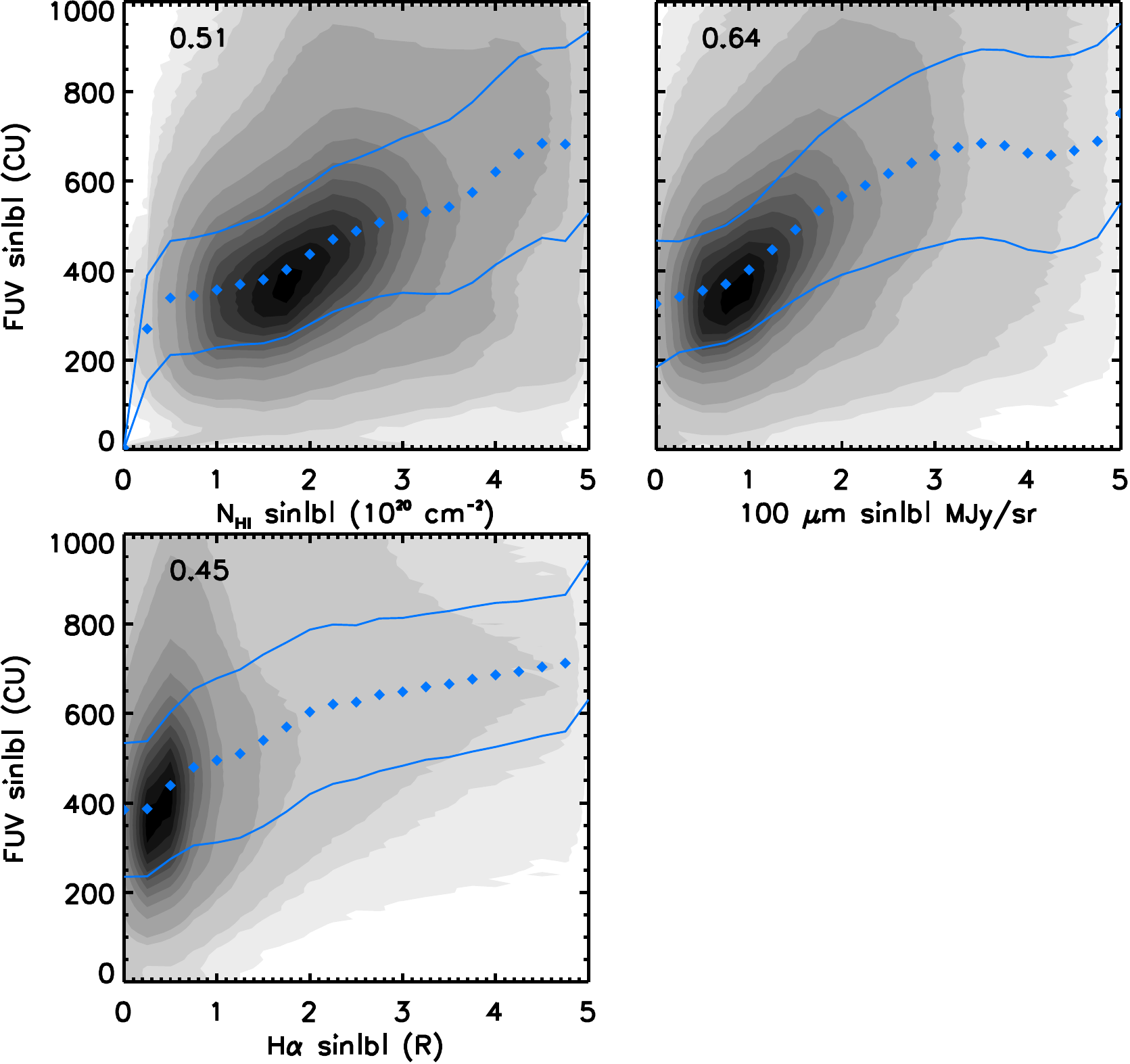}
\caption{2-d histograms of FUV sin$|$b$|$ vs. Galactic quantities times sin$|$b$|$.  Blue dots indicate the median value for abscissa bins of 0.25, while blue lines indicate one standard deviation from the median.  Linear Pearson correlation coefficients are shown in the upper left of each panel.  \textbf{Top Left:} FUV sin$|$b$|$ vs. N$_{\rm HI}$ sin$|$b$|$. \textbf{Top Right:} FUV sin$|$b$|$ vs. 100 $\mu$m sin$|$b$|$.  \textbf{Bottom Left:} FUV sin$|$b$|$ vs. H$\alpha$ sin$|$b$|$.}\label{fig:fuvvsglobalsinb}
\end{figure}

As noted above, any quantity which has a plane parallel distribution with respect to the Galactic plane will vary with latitude roughly as the cosecant of latitude.  By removing the cosecant dependence, we can verify that deviations from a plane parallel distribution are also correlated between two different Galactic quantities.  As such, we re-plot Figure \ref{fig:fuvvsglobal} with a factor of sin$|$b$|$, shown in Figure \ref{fig:fuvvsglobalsinb}.  The correlation coefficient is again calculated using the linear Pearson method and are weaker after the cosecant correction.  The scatter in all plots is increased compared to Figure \ref{fig:fuvvsglobal}.  The correlation between 100 $\mu$m emission and diffuse FUV remains the strongest and is discussed further in Section \ref{sec:slopes}.  The correlation between H$\alpha$ and diffuse FUV is still the weakest of the three, and we examine it in more detail in Section \ref{sec:fuvhalpha}.  \citet{2011Seonb} noted that 100 $\mu$m emission and N$_{\rm HI}$ corresponded well to a plane parallel model, while H$\alpha$ and FUV emission did not.  As a simple plane parallel model only crudely represents the true 3-D distribution of any Galactic component (\citealt{1994Witt} and discussed above), it is not surprising that we find these weak correlations. A more detailed model is required to fully interpret this result.

\begin{figure}
\centering
\includegraphics[width=0.45\textwidth]{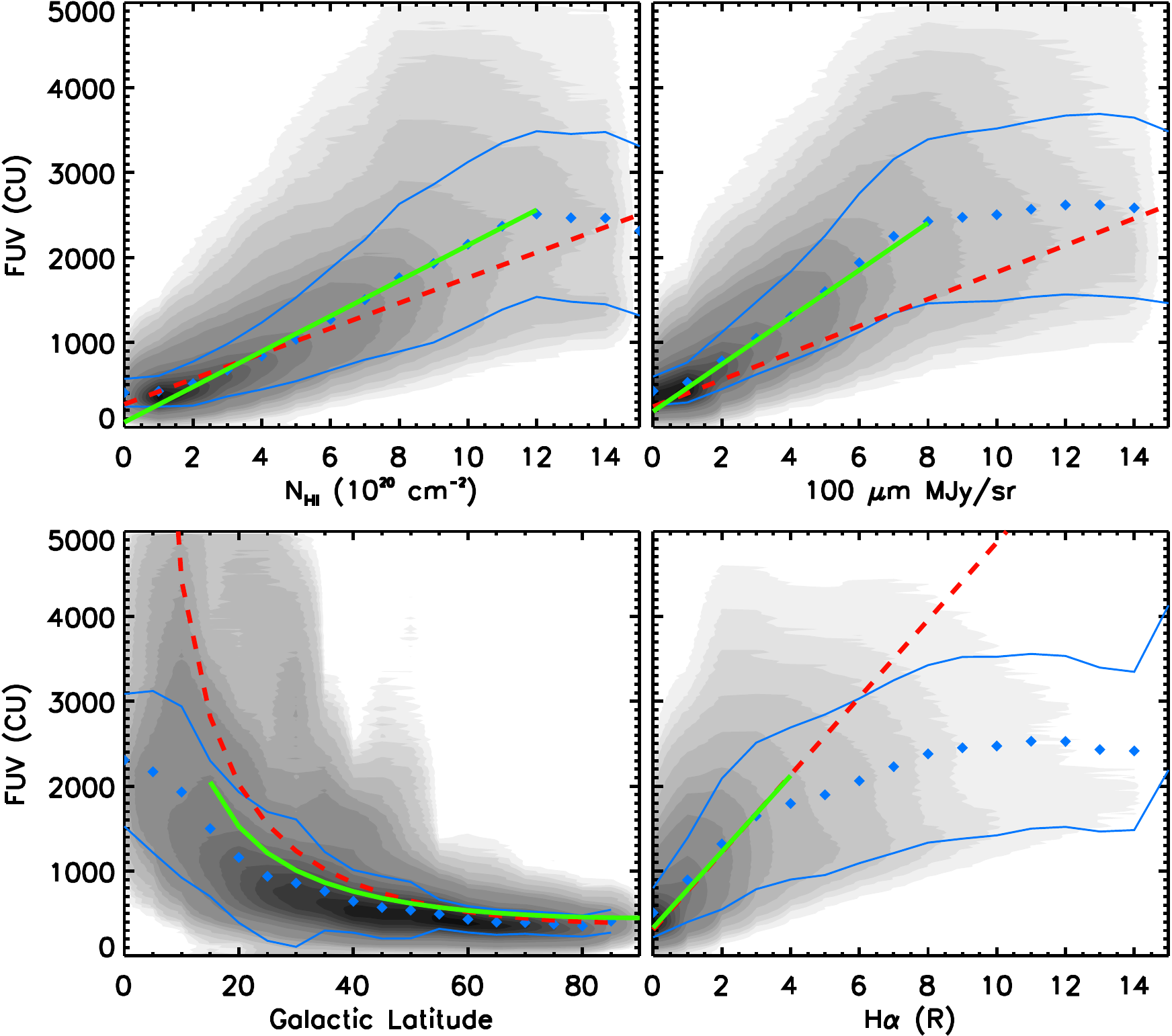}
\caption{2-D histograms of FUV vs. Galactic quantities with a linear scale.  Blue points indicate the median for abscissa bins of 1.0, while the blue lines indicate one standard deviation from the median.    Green lines are best fit lines for the median.  Red lines are best fits lines from \citet{2011Seonb}. \textbf{Top Left} FUV vs. N$_{\rm HI}$.  The line fit is restricted to 0-1.2 $\times$ 10$^{21}$ cm$^{-2}$, above which there appears to be a flattening off. \textbf{Top Right:} FUV vs. 100 $\mu$m.  The line fit is restricted to 0-8 MJy/sr.  \textbf{Bottom Left:}  FUV vs. sin $|$b$|$.  The fit to the median is a two parameter cosecant function, I=A/sin$|$b$|$+D for b $>$ 25$^\circ$.  Here A=457.9$\pm$248.7, D=-88.5$\pm$283.0.  \textbf{Bottom Right:} FUV vs. H$\alpha$.}\label{fig:fuvvsglobal_smooth}
\end{figure}

A comparison of the quantities is shown in Figure \ref{fig:fuvvsglobal_smooth}, to highlight behavior at low intensities where the relationship is primarily linear.  \citet{2011Seonb} calculated best fit lines (in red) for b$>$25$^\circ$, while our lines include data from all latitudes.  The fits are similar for all but 100 $\mu$m.  Restricting our data to b$>$25$^\circ$ yields a closer match in fit for FUV vs. 100 $\mu$m.  Table \ref{table:bestfits} shows slopes and intercepts for the calculated best fit lines.  

Because the correlation with dust (and other properties) is nearly linear at low intensities, it is conventional to use the fit to this relation to determine the value of the constant FUV offset, which presumably includes components that are not associated with dust-scattered light.   Most analyses have assumed that this component is nearly isotropic and we do the same here.  FUV vs 100 $\mu$m emission shows a pronounced flattening at low 100 $\mu$m values ($<$ 1 MJy/sr, as seen in the log-log plot of Figure \ref{fig:fuvvsglobal}).  Furthermore, there is a FUV offset in the plots in Figure \ref{fig:fuvvsglobal_smooth}, at zero values of the abscissa.  This minimum appears to be 200-300 CU.  This offset has also been noted as a positive offset at N$_{\rm HI}$=0 cm$^{-2}$ of 200-300 CU by \citet{1991Martin}, who suggest that it may be partially due to an undetected dust component.  These offsets have been discussed in other works as a combination of a low level extragalactic FUV background (a few tens of CU, \citealt{1990Paresce,2001Schiminovich}), incomplete bright object masking and airglow contamination.  In this work, we have used 300 CU as the offset and revisit this component in Section \ref{sec:conclusion}.

\begin{table*}
\normalsize
\centering
\caption{Best fit lines- slopes and intercepts}
\label{table:bestfits}
\begin{tabular}{|c|c|c|c|c|c|c|c|c|}
\hline
& & all points & 0-15$^\circ$ & 15-30$^\circ$ & 30-45$^\circ$ & 45-60$^\circ$ & 60-75$^\circ$ & 75-90$^\circ$  \\
\hline
100 $\mu$m$^a$ & slope & 275$\pm$95 & 253$\pm$68 & 250$\pm$53 & 168$\pm$43.9 & 100$\pm$33 & 62$\pm$18 & 82$\pm$67 \\
 & intercept & 241$\pm$273 & 601$\pm$201 & 411$\pm$139 & 480$\pm$109 & 432$\pm$95 & 386$\pm$73 & 338$\pm$98 \\
\hline
N$_{\rm HI}$$^b$ & slope & 184$\pm$50 & 223$\pm$50& 123$\pm$39& 121$\pm$30& 88$\pm$18& 68$\pm$21 & 6.3$\pm$33\\
& intercept & 188$\pm$180 & 123$\pm$240 & 636$\pm$169& 403$\pm$117 & 347$\pm$76& 305$\pm$86 &,383$\pm$99\\
 \hline
H$\alpha$$^c$ & slope & 383$\pm$478 & 214$\pm$61 & 161$\pm$58 & 128$\pm$44 & 83$\pm$19 &7$\pm$42 & 39$\pm$105 \\
&intercept & 519$\pm$859 & 1342$\pm$281 & 938$\pm$269 & 680$\pm$189 & 547$\pm$112 & 434$\pm$96 & 379$\pm$108 \\
\hline
\multicolumn{7}{l}{\small ~$^a$~ $<$ 8 MJy/sr for 100 $\mu$m}\\
\multicolumn{7}{l}{\small ~$^b$~ $<$ 10$^{21}$ cm$^{-2}$ for N$_{\rm HI}$}\\
\multicolumn{7}{l}{\small ~$^c$~ $<$ 4 R for all points only, $<$ 10 R for all others, for H$\alpha$}\\
\end{tabular}

\end{table*}

In the top right panel of Figure \ref{fig:fuvvsglobal_smooth}, we note the break in the FUV intensity at 100 $\mu$m $>$ 8 MJy/sr, with a median FUV intensity of 2400 CU.  This saturation in the FUV intensity has been noted previously \citep{2011Seonb,2008Witt}, and appears to occur for lines of sight having an optical depth high enough to both self-shield emission from within the cloud and block scattered FUV intensity from behind the cloud, decreasing the overall FUV intensity from that region. At high 100 $\mu$m we also observe a large scatter in FUV intensity.  Along some sightlines, the presence of nearby FUV bright stars can enhance the overall FUV intensity above that predicted under the assumption of a uniform radiation field.  A more detailed analysis of UV self-shielding and illumination of these sightlines is discussed in a forthcoming paper.  

We find a similar break in the correlation of FUV vs. N$_{\rm HI}$ at N$_{HI} \sim$12 $\times$ 10$^{21}$ cm$^{-2}$.  There is a break in the plot of FUV intensity vs. H$\alpha$ intensity at 4 R, as found by \citet{2011Seonb}, although this is a more gradual transition than for the other two quantities.  In all plots, there is increased scatter as the abscissa values increase.  If the FUV intensity is log normal, as described in \citet{2013Seon}, this is a natural property of log normal distributions.  However, the plot of FUV vs. Galactic latitude shows this occurs primarily at the lowest latitudes.  Cutting out intensity from points with $|$b$|$ $<$ 25$^\circ$ decreases this scatter and it's likely the strong ISRF at low latitudes contributes significantly to the scatter.  

\begin{figure}
\centering
\includegraphics[width=0.45\textwidth]{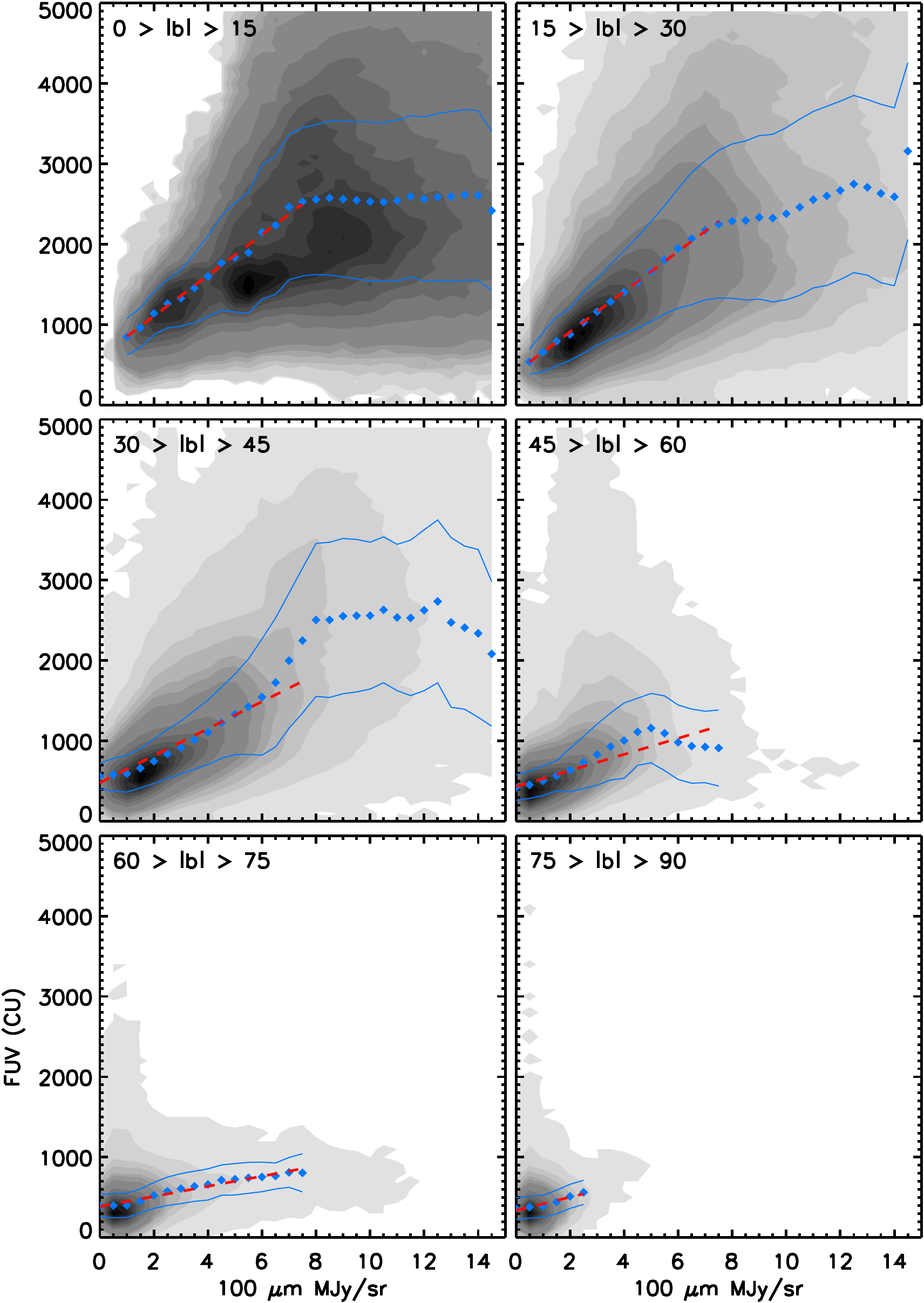}
\caption{2-D histograms of FUV intensity vs. 100 $\mu$m emission for latitude cuts of 15$^\circ$, N and S combined.  Blue dots are the median for bins of 0.5 MJy/sr, with blue lines indicating one standard deviation.  The red line is the best fit line to the median below 8 MJy/sr.  At low latitudes, there is significant scatter in the relationship, due to both FUV bright stars and obscuring dust.  There is a turnover in FUV intensity at $\sim$8 MJy/sr, above which the median FUV value remains constant.  The slope of the linear relationship decreases systematically with increasing latitude, becoming smaller at the highest latitude cut.}\label{fig:irabslat}
\end{figure}

\subsection{Correlations vs. Galactic latitude}

We have already noted above that a difficulty in comparing our results to that of previous work is that derived correlations will depend on the Galactic footprint of the data used, and in particular the range in latitude.  In order to understand the magnitude and physical origin of these effects, it is useful to divide our large data set into Galactic latitude cuts.  In doing so, we note that regions of high scatter are generally confined to the lowest latitudes, while the FUV emission adheres to a linear fit at higher latitudes.

Figure \ref{fig:irabslat} shows contour plots of FUV vs. 100 $\mu$m for latitude bins of 15$^\circ$, combining the northern and southern hemispheres.  Above 8 MJy/sr in latitude bins from 0-15$^\circ$ and from 30-45$^\circ$, there is a flattening in the FUV profile, with the median constant at around 2500 CU.  This flattening is not as clear in the latitude cut from 15-30$^\circ$, but this is likely due to very high FUV emission around the Ophiuchus and Orion OB associations. 

At high latitudes $|$b$| >$ 45$^\circ$, few points have 100 $\mu$m values above 8 MJy/sr, leaving the low intensity linear relationship.  The slope of the linear portion does appear to change with latitude, with the low latitude cuts having a larger slope than at higher latitudes.  Line fits below 8 MJy/sr are plotted in red, with fits as in Table \ref{table:bestfits}. 

\begin{figure}
\centering
\includegraphics[width=0.45\textwidth]{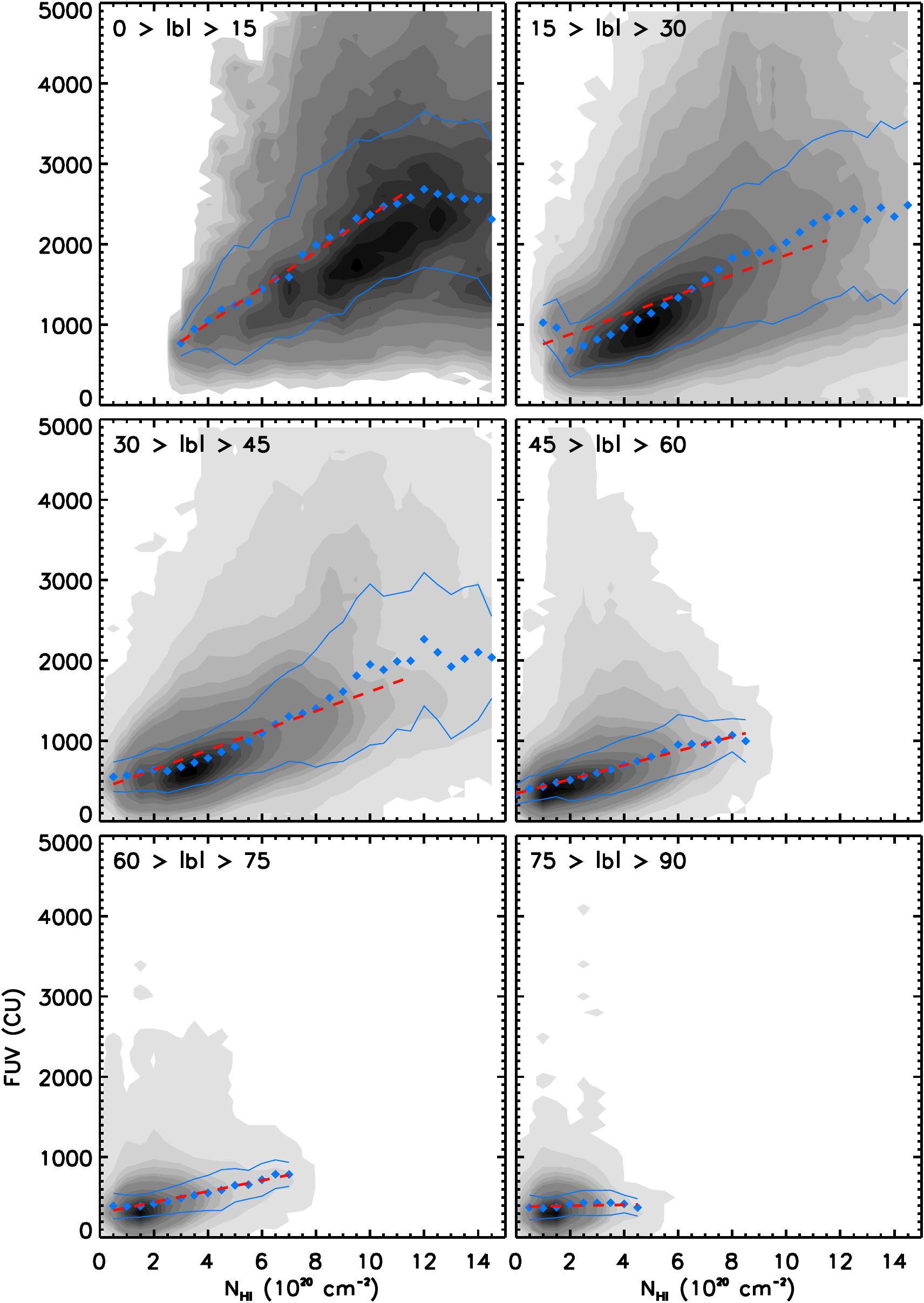}
\caption{2-D histogram of FUV intensity vs. N$_{\rm HI}$ column.  Blue and red lines as in Figure \ref{fig:irabslat}.  There is a turnover in FUV intensity at $\sim$10$^{21}$ cm$^{-2}$.}\label{fig:h1abslat}
\end{figure}

\begin{figure}
\centering
\includegraphics[width=0.45\textwidth]{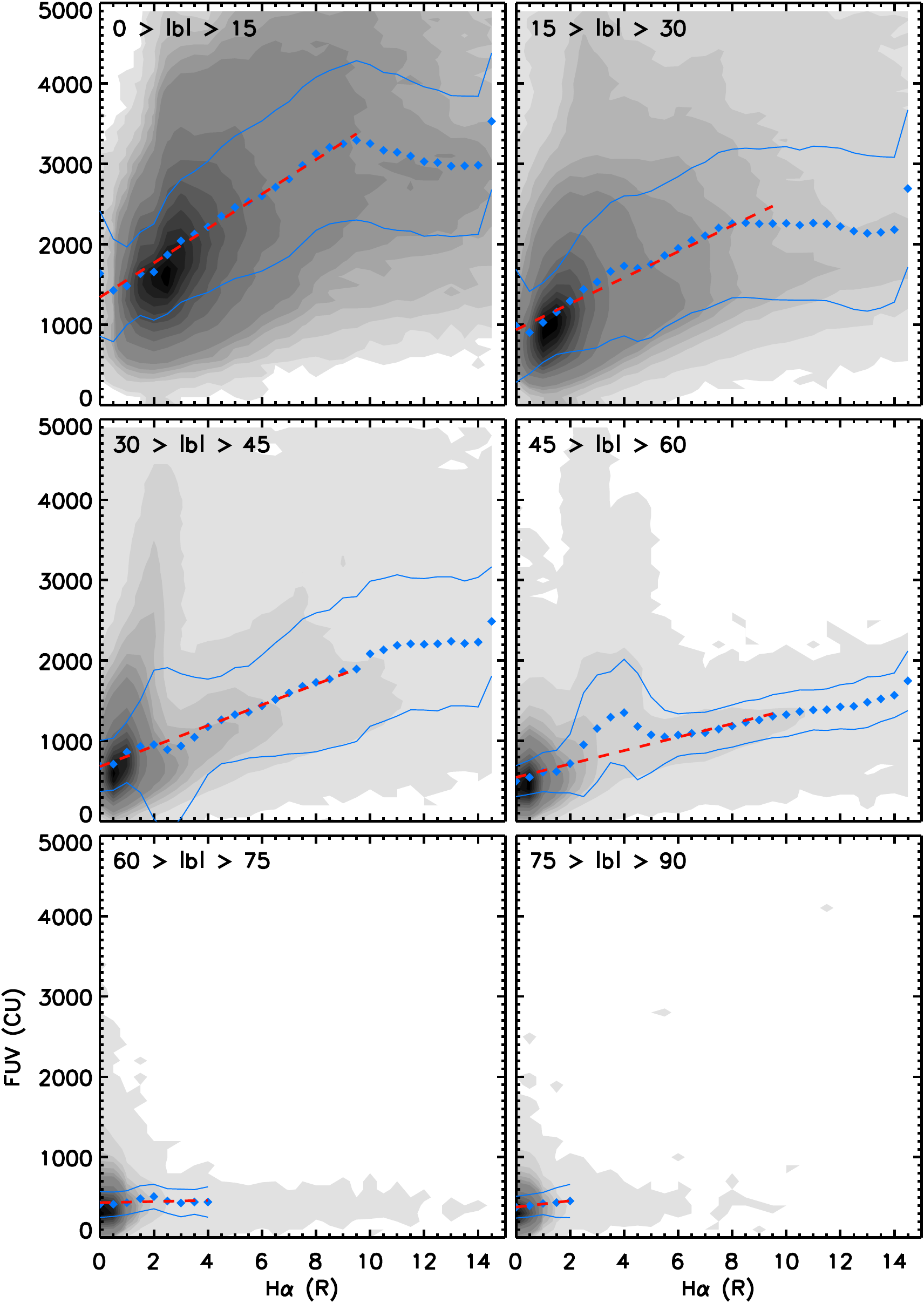}
\caption{FUV intensity vs. H$\alpha$ intensity. Blue and red lines as in Figure \ref{fig:irabslat}.}\label{fig:halphaabslat}
\end{figure}

Some of the scatter from the linear relation can be directly traced to specific objects or regions in the Galaxy.  For example, in the latitude cut 45-60$^\circ$ there is a region of low 100 $\mu$m, but very high FUV ($>$ 2500 CU), which stands out compared to the rest of the high latitude region.  This appears to come primarily from the region directly around Spica ($\alpha$ Vir, at $l$ = 316$^\circ$, $b$ = 51$^\circ$, \citealt{2012Park}), which is a spectroscopic binary with two B type stars.  The FUV intensity here is high, while the dust emission is more consistent with the rest of the latitude.  This area also appears in Figures \ref{fig:h1abslat} and \ref{fig:halphaabslat}, at similarly low values for the abscissa.

At very high latitudes, where there appears to be only a weak relationship between dust and FUV intensity, increased relative scatter may be masking any correlation.  This is reflected in the Pearson r value for the fits which is generally observed to decrease with latitude.  For example, while the overall correlation between FUV and 100 $\mu$m is quite high (r = .80 for all points in log-log space, r = .64 after removing the cosecant dependence), at high latitudes, the linear Pearson r value drops to 0.14 (for $|$b$| >$ 75$^\circ$, with or without the cosecant dependence).  The FUV intensity in these regions is quite low, and the scatter is relatively large enough to give the appearance of high latitude FUV intensity that is only weakly sensitive to the 100 $\mu$m emission.  

Figure \ref{fig:h1abslat} shows contour plots of FUV vs. N$_{\rm HI}$ column for $|$b$|$ cuts of 15$^\circ$.  As with Figure \ref{fig:irabslat}, there is a flattening of the FUV emission at high values of N$_{\rm HI}$.  This appears to occur at 12 $\times $10$^{20}$ cm$^{-2}$, which is consistent with the behavior found by \citet{1991Hurwitz} and others more recently including \citet{2011Seonb}.  Low column at high latitudes means that this flattening column density is not reached above $|$b$|$ = 45$^\circ$.  As with Figure \ref{fig:irabslat}, the slope of the linear portion changes between latitude cuts.  Line fits below 10 $\times $10$^{20}$ cm$^{-2}$ are plotted in red, with fits as in Table \ref{table:bestfits}. 

Figure \ref{fig:halphaabslat} shows contour plots of FUV vs. H$\alpha$ intensity for $|$b$|$ cuts of 15$^\circ$.  Here, there is much more scatter than with 100 $\mu$m or N$_{\rm HI}$.  We also observe variation in the H$\alpha$ intensity where FUV plateaus.  In the low latitude cuts, the turnover appears at 8-10 R, while at mid latitudes (30$ < |$b$| < $60), there is little evidence for a turnover even above these H$\alpha$ values.  At the highest latitudes ($|$b$| >$ 60) the relationship is nearly flat.  Line fits below 10 R are plotted in red, with fits as in Table \ref{table:bestfits}.  Along with the high FUV intensity region around Spica mentioned above, there is a region of high FUV intensity at latitudes between 30-60$^\circ$.  These can be traced to the region directly above the Ophiuchus association ($l$=355$^\circ$, $b$=18$^\circ$).  This region has significantly more scattered FUV than H$\alpha$.  The Spica and Ophiuchus regions cause a bump in the median FUV value between 1-4 R.  A more detailed discussion of FUV vs. H$\alpha$ is found in Section \ref{sec:fuvhalpha}.  

\subsection{FUV vs. 100 $\mu$m}
\label{sec:slopes}

\begin{figure}
\centering
\includegraphics[width=0.5\textwidth]{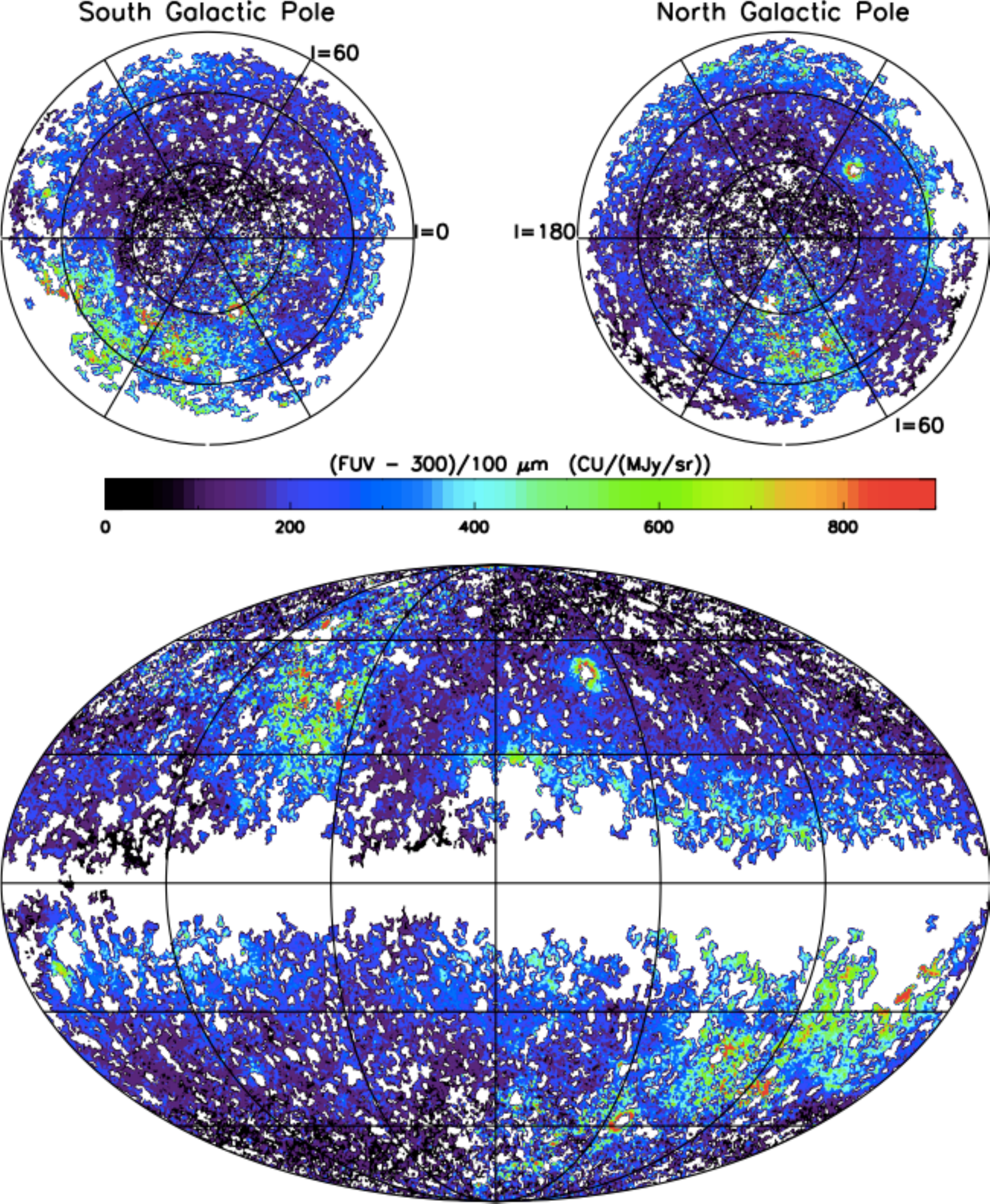}
\caption{FUV/100 $\mu$m ratio (CU/(MJy/sr)) for the whole sky.  The FUV offset of 300 CU has been removed for these plots.  Notable high slope regions are around OB associations, e.g. Ophiuchus, and brights stars, e.g. the spectroscopic binary Spica (at l=316, b=51., \citet{2012Park}).}\label{fig:allskyfuvir}
\end{figure}

The relationship between diffuse FUV intensity and 100 $\mu$m emission is often expressed as a slope with units of CU/(MJy/sr).  Previous work shows a wide range of slopes.  \citet{1991Perault} measured 244 CU/(MJy/sr) in the northern hemisphere and 214 CU/(MJy/sr) in the southern hemisphere, using data from the ELZ spectrophotometer on DB2-AURA.  \citet{1991Hurwitz} found  $\sim$294 CU/(MJy/sr) using data from the Berkeley spectrometer on UVX.  \citet{1992Wright} obtained 203 CU/(MJy/sr), using data from \citet{1989Fix}. \citet{1995Haikala} used FAUST data to observe Galactic cirrus near the north Galactic pole, finding 128 CU/(MJy/sr).   \citet{1996Sasseen} found a range of slopes from -49 to 255 CU/(MJy/sr) in 13 regions using data from FAUST.  \citet{2010Sujatha} find slopes between 50 and 480 CU/(MJy/sr) using data from GALEX of part of the Draco Nebula.  \citet{2010Murthy} find an average slope of 302 CU/(MJy/sr) using smoothed GALEX data from the whole sky. \citet{2011Seonb} find a slope of 158 CU/(MJy/sr) from SPEAR data.  In our all sky GALEX data, we find an average slope of 280 CU/(MJy/sr).  

Clearly, the behavior of FUV intensity and 100 $\mu$m emission can vary significantly from region to region, and even within the same cloud complex.  We show a contour plot of ratios (FUV/IR) for the whole sky in Figure \ref{fig:allskyfuvir}.  The isotropic diffuse FUV offset of 300 CU is removed.  There are two large regions with high ratios, one in the northern hemisphere above $b>$30$^\circ$ between $l$=60$^\circ$ to $l$=180$^\circ$ and one in the southern hemisphere below $b<$-45$^\circ$ between $l$=240$^\circ$ to $l$=0$^\circ$.  These features correspond to regions of particularly low IR emission, with 100 $\mu$m intensity of less than 1 MJy/sr, and often lower than 0.5 MJy/sr.  In their original dust map, \citet{1998Schlegel} note these extremely low emission windows as good regions for observations requiring minimum dust contamination.  

Very low ratios ($<$ 100 CU/(MJy/sr)) are found near the Galactic plane, reflecting the high dust content in these regions.  There are some low latitude areas with high ratios from excess FUV intensity due to proximity of nearby OB associations, particularly Orion ($l$=200$^\circ$), the Gum nebula ($l$=260$^\circ$,$b$=-2$^\circ$), and Ophiuchus($l$=355$^\circ$, $b$=18$^\circ$).  

\begin{figure}
\centering
\includegraphics[width=0.45\textwidth]{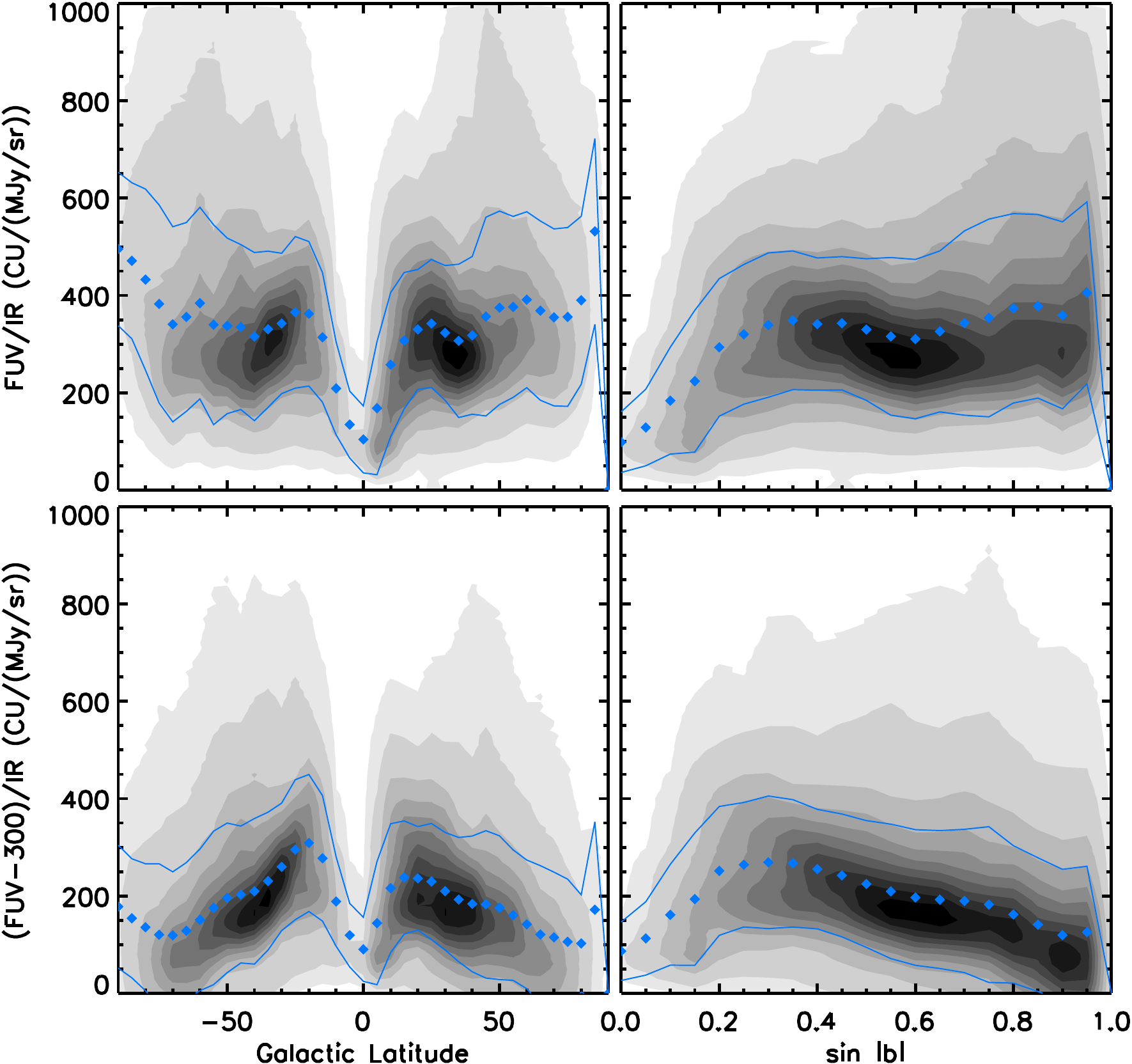}
\caption{2-D histograms of FUV/IR (CU/(MJy/sr) vs. Galactic coordinates.  Blue dots indicate the median for bins of 5$^\circ$, with blue lines indicating the standard deviation from the median. \textbf{Top Left:} FUV/IR ratio vs. Galactic latitude.  \textbf{Top Right:} The same data as in the top left plot, but plotted vs. sin$|$b$|$.  \textbf{Bottom Left:} Same as above, but with the 300 CU offset removed from the FUV data.  \textbf{Bottom Right:} Same as above, but with the 300 CU offset removed from the FUV data.  The declining slope above sin$|$b$|$ = .4 and the decrease in FUV/IR ratio with increasing latitude both suggest a deviation from a simple plane parallel distribution. }\label{fig:slope_lat}
\end{figure}

In Figure \ref{fig:slope_lat}, showing the FUV/IR ratio vs. Galactic latitude, the ratio is nearly constant.  With the 300 CU offset removed, as in the bottom panels, the ratio begins to decline at latitudes above $|$b$|$=30$^\circ$, the same behavior as the slopes in Figure \ref{fig:irabslat}.  This is likely driven by decreasing FUV intensity at high latitudes, rather than increasing IR intensity.  In both panels, there is significant scatter at all latitudes.  

\begin{figure}
\centering
\includegraphics[width=0.45\textwidth]{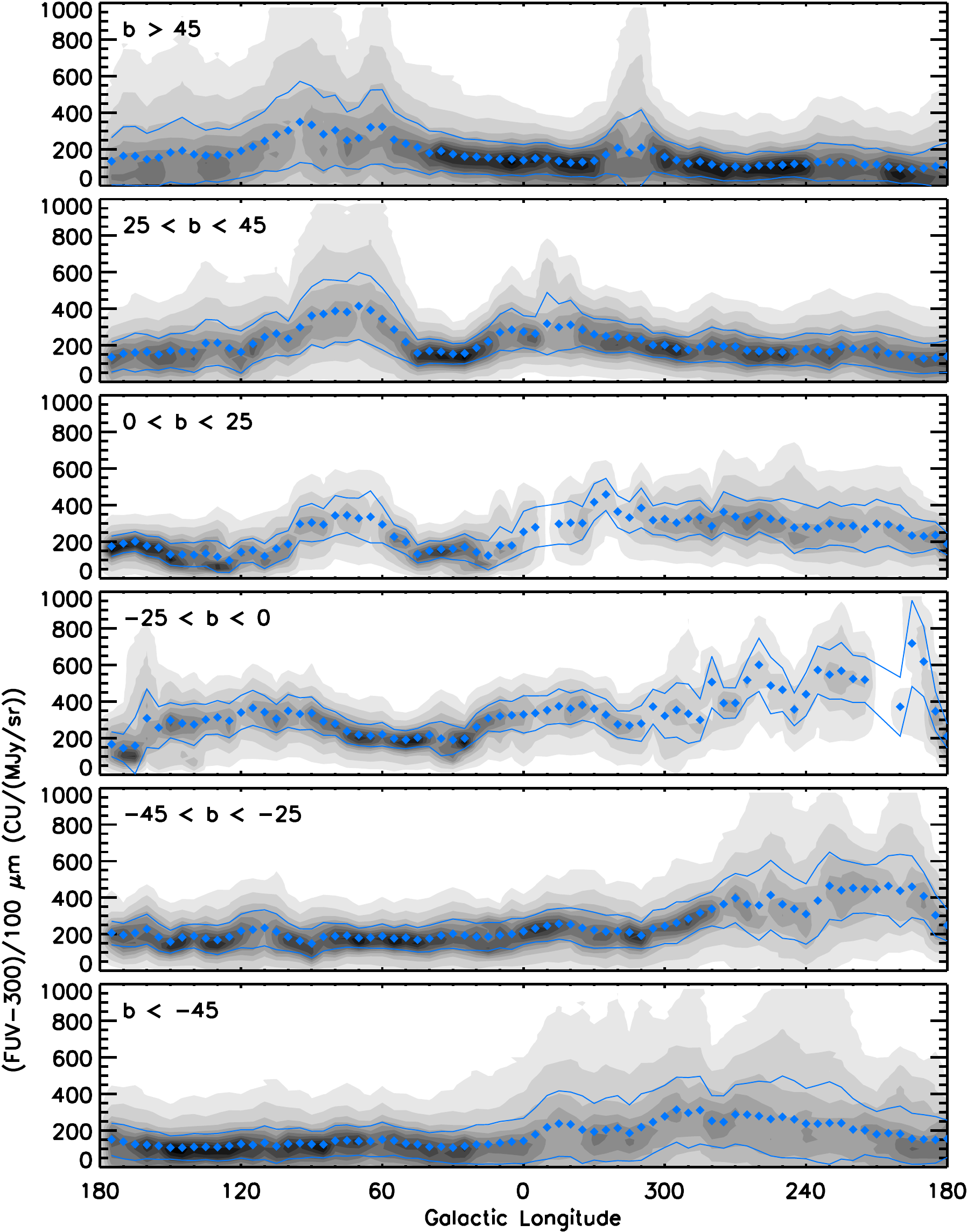}
\caption{All plots show 2-D histograms of FUV/IR (CU/(MJy/sr)) vs. longitude for latitude cuts as indicated, with the 300 CU offset removed from the FUV data.  Blue dots indicate the median for bins of 5$^\circ$ with blue lines indicating one standard deviation.  These plots show both the median slope, 280 CU/(MJy/sr), but also the large regional variations.  Some variations are spatially coherent, as evidenced by high slopes at northern latitudes around $l$=60$^\circ$ and southern latitudes around $l$=260$^\circ$.}\label{fig:slope_long}
\end{figure}

The origin of this scatter becomes more clear in Figure \ref{fig:slope_long}, which shows the ratio of FUV/IR vs. Galactic longitude for different latitude cuts.  The high ratio regions centered around $l$=90$^\circ$ in the northern hemisphere and at $b <$ -45$^\circ$, 350 $< l <$ 250$^\circ$ in the southern hemisphere, are the same regions that have been noted previously.  Otherwise, elevated ratios at low latitudes indicate higher than expected FUV intensity.  In particular, the Orion OB complex, the Gum nebula, and regions in between have excess FUV intensity, concentrated in the Galactic plane.  There is little leakage of this to higher latitudes, as evidenced by the general flat profile in the top two panels of Figure \ref{fig:slope_long}.  High slopes in the southern hemisphere below this region could point to leakage of FUV photons, but that may also be the result of the low IR emission region discussed above.  The region of high slopes near b$>$ 45$^\circ$ at $l$=310$^\circ$ is again due to excess FUV intensity from Spica.

\subsection{FUV vs. H$\alpha$}
\label{sec:fuvhalpha}

The relationship between diffuse FUV intensity and diffuse H$\alpha$ intensity has not been as well studied as other Galactic quantities.  For high latitude H$\alpha$ intensity in particular (tracing the diffuse warm ionized medium, WIM), the common assumption was that most Galactic H$\alpha$ originated in ionized HII regions, with significant leakage of Lyman continuum photons responsible for the H$\alpha$ intensity observed elsewhere \citep{1995Reynolds,2009Haffner}, although the mechanism to provide the necessary leakage has not been convincingly explained \citep{2009Seon,2010Wood,2012Seon}.  Recent work by \citet{2010Witt}, \citet{2011Seona}, and \citet{2011Dong} have argued that a significant percent of H$\alpha$ intensity observed outside of HII regions is in fact dust scattered, and can be shown to correlate with the diffuse FUV intensity.  

Scattering percentages for H$\alpha$ in the WIM have been calculated to be as low as 5-20\% \citep{1999Wood}, 20\% \citep{2011Dong}, and as high as 37\% \citep{2011Seona}, using varied techniques.  \citet{2012Brandt} have recently measured the visible spectrum of diffuse Galactic light, and in so doing find that scattering accounts for around 19\% $\pm$ 4\% of H$\alpha$ intensity, for $|$b$| >$ 60$^\circ$.  

As shown in Figures \ref{fig:fuvvsglobalsinb} and \ref{fig:fuvvsglobal}, there is a correlation between the diffuse FUV and H$\alpha$ intensity, although it is not as tightly correlated as 100 $\mu$m and N$_{\rm HI}$, with r=0.73 (log-log) overall and r=0.45 after the latitude dependence is removed.  Still, this indicates that there is some shared dependence between FUV and H$\alpha$ as discussed above.  Our data set mainly encompasses the diffuse WIM due to the avoidance of the Galactic plane and bright regions.

\begin{figure}
\centering
\includegraphics[width=0.45\textwidth]{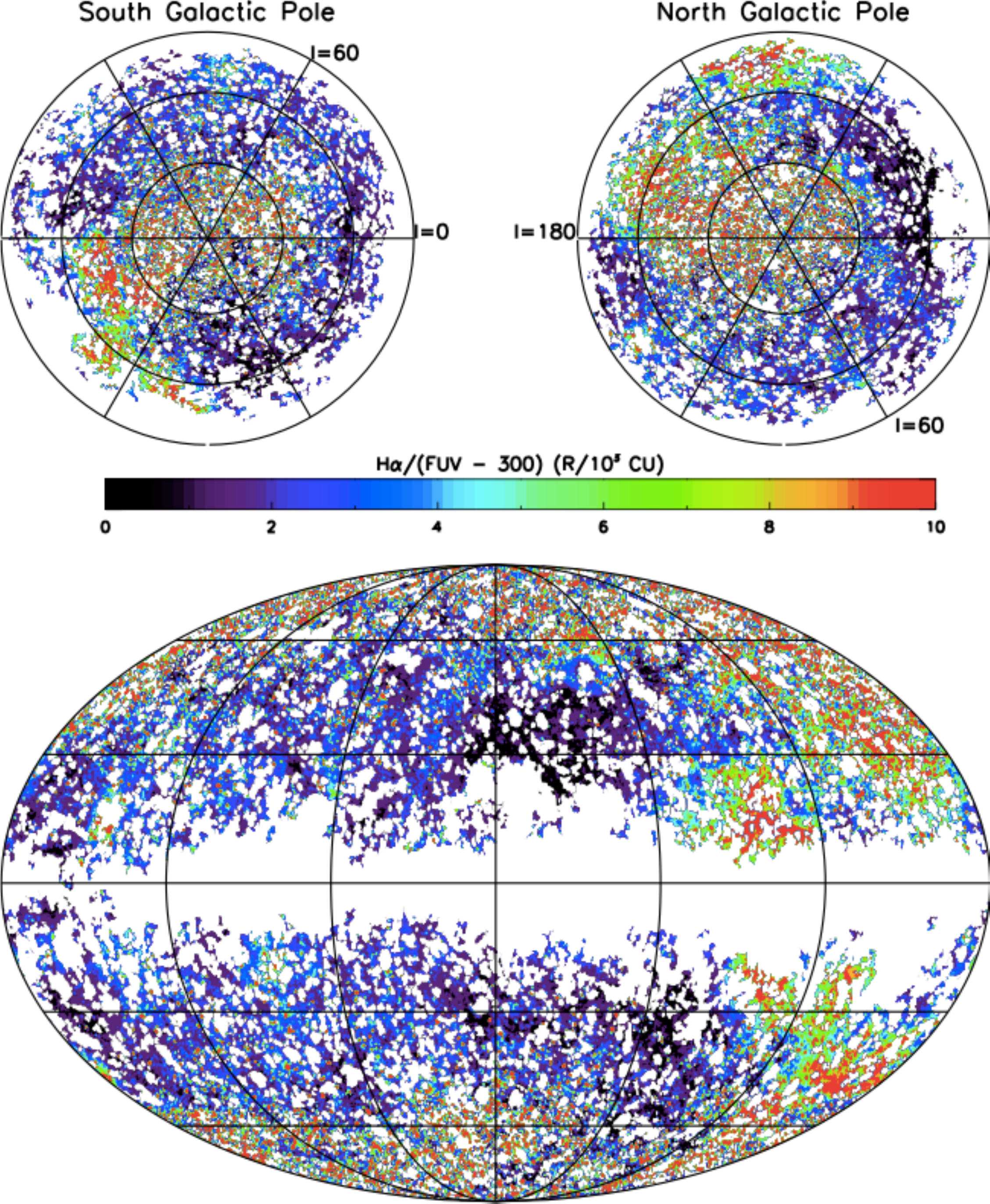}
\caption{All-sky map of H$\alpha$/FUV, in units of R/10$^3$ CU.  A 300 CU offset for the FUV map has been removed.}\label{fig:halphafuvallsky300}
\end{figure}

Figure \ref{fig:halphafuvallsky300} shows an all-sky map of H$\alpha$/FUV in units of R/10$^3$ CU, following \citet{2011Seona}.  An offset of 300 CU has been subtracted from the FUV data.  There is clear structure, including especially high ratios around the Gum Nebula and Orion complex.  These are all likely due to high H$\alpha$ intensity from HII regions.  Other OB associations, including Ophiuchus and structures from $l$=0 to 180$^\circ$, do not have the same high ratios, despite similar FUV emission values.  Finally, there are high ratios at both poles, primarily driven by low FUV intensity than high H$\alpha$.  Outside of these regions, at mid-latitudes, there is a relatively stable ratio of 2-3 R/10$^3$ CU.

\begin{figure}
\centering
\includegraphics[width=0.45\textwidth]{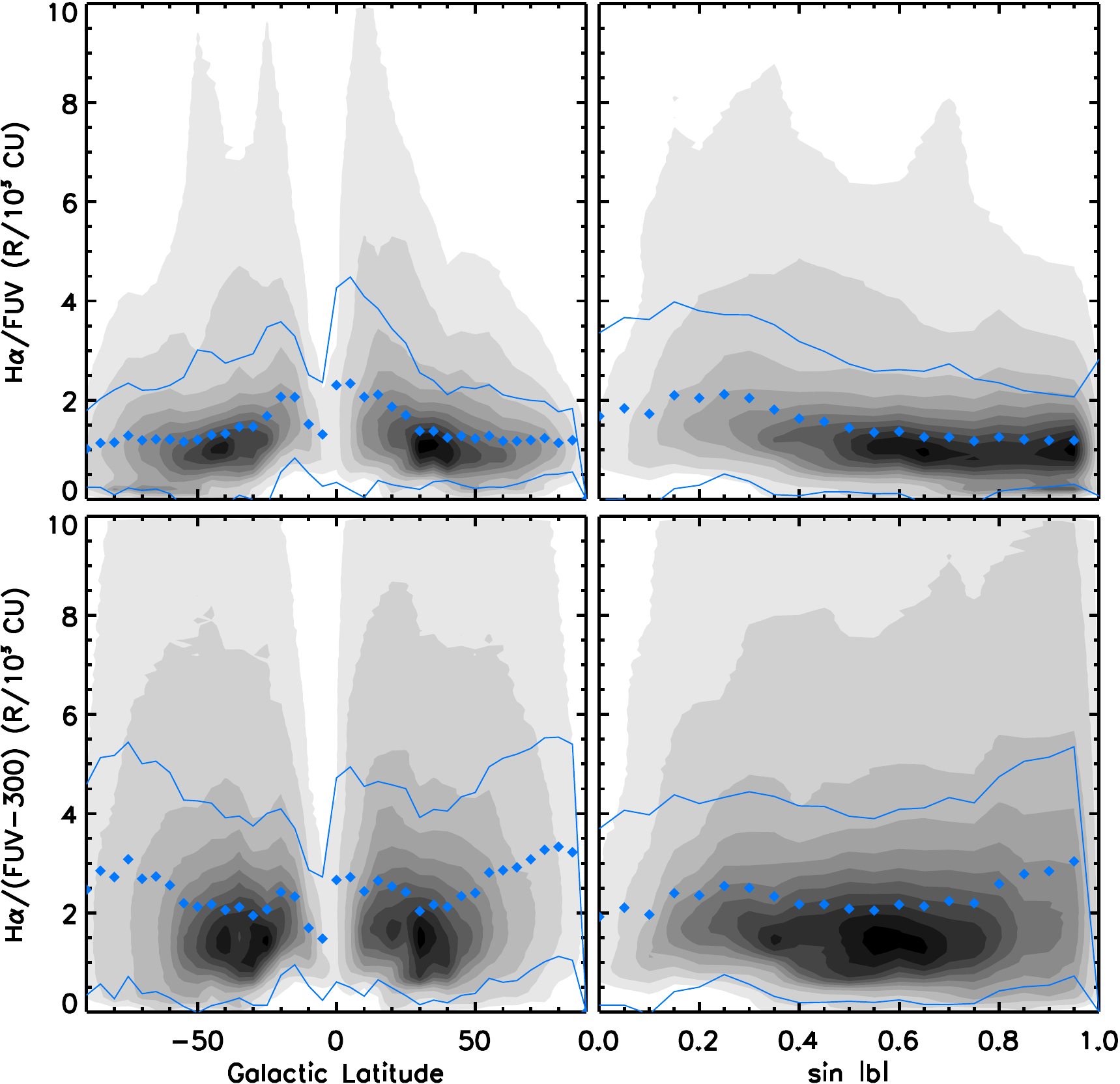}
\caption{2-D histograms of H$\alpha$/FUV (R/10$^3$ CU) vs. Galactic coordinates.  Blue dots indicate the median for latitude bins of 5$^\circ$ with blue lines indicating one standard deviation.  \textbf{Top Left:} A 2-d histogram of H$\alpha$/FUV vs. Galactic latitude.  \textbf{Top Right:} The same data plotted vs. sin$|$b$|$.  \textbf{Bottom Left:} Same as above, but with a 300 CU offset removed from the FUV data.  \textbf{Bottom Right:} Same as above, but with a 300 CU offset removed from the FUV data.}.\label{fig:slopelathalpha}
\end{figure}

Figure \ref{fig:slopelathalpha} shows the ratio of H$\alpha$/FUV as a function of latitude.  The higher ratios at low latitudes are the result of bright HII regions, but in general the ratio is nearly constant.  Unlike for FUV/100 $\mu$m (Figure \ref{fig:slope_lat}), the plot of H$\alpha$/FUV vs. Galactic latitude doesn't appear significantly changed by the removal of the FUV offset.  Potentially this is because the H$\alpha$ intensity is the result of a wide range of processes, not just scattering, so the correlation is low to begin with (as noted in Figure \ref{fig:fuvvsglobalsinb}).  The range of ratios becomes larger at high latitudes, but the median remains roughly constant below sin$|$b$|$=0.8, and rises slightly after.

\begin{figure}
\centering
\includegraphics[width=0.45\textwidth]{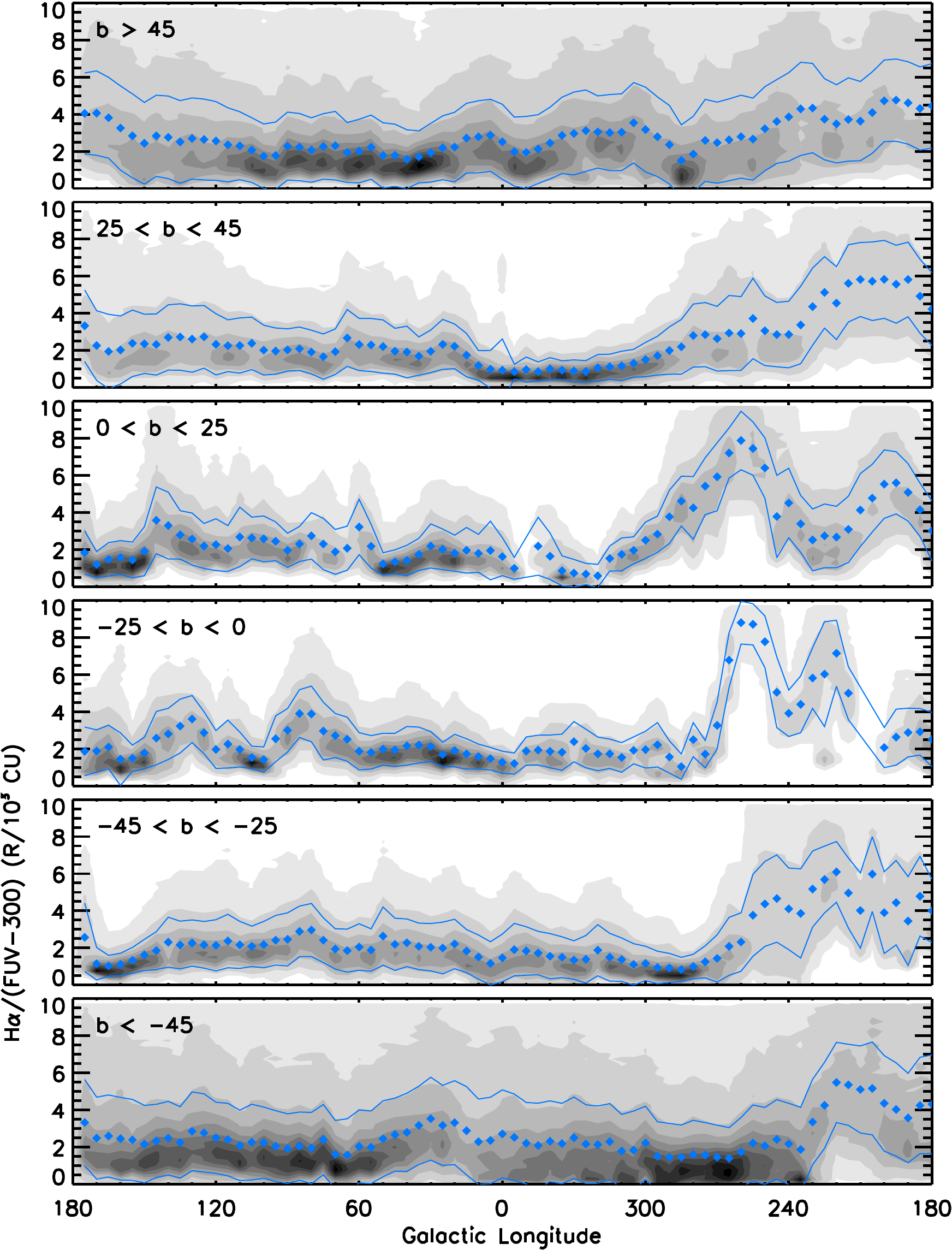}
\caption{All plots show 2-D histograms of H$\alpha$/FUV (R/10$^3$ CU) vs. longitude for latitude cuts as indicated, with the 300 CU offset removed from the FUV data. Blue dots indicate the median for longitude bins of 5$^\circ$ while blue lines indicate one standard deviation.  These plots show both the low level average slope (2-4 R/10$^3$ CU), but also the large regional variations.  Some variations are spatially coherent, as evidenced by high slopes at southern latitudes around $l$=260$^\circ$ and 200$^\circ$.}\label{fig:fuvHalpha_latcuts}
\end{figure}

Figure \ref{fig:fuvHalpha_latcuts} shows the ratio of H$\alpha$/FUV as a function of Galactic longitude for different latitude cuts, with the 300 CU FUV offset removed.  Like its counterpart for 100 $\mu$m emission Figure \ref{fig:slope_long}, the ratio varies by an order of magnitude across the sky.  At the highest latitude cuts, the standard deviation is 2 R/10$^3$ CU, but the mean is relatively stable with longitude.  At latitudes closer to the Galactic plane, the standard deviation decreases, but the variation in ratio can be more than a factor of 2 between different longitudes.  Some of this variation is seen in multiple latitude cuts, with high H$\alpha$/FUV ratios appearing in the same longitude range.  Of particular note is the peak at $l$=200$^\circ$, potentially associated with the Orion OB association, which appears at all latitude cuts.  This peak is the result of high H$\alpha$ intensity and recalls a similarly placed peak in Figure \ref{fig:slope_long}.  In some cases, excess H$\alpha$ intensity may be caused by significant Lyman continuum photon leakage into high latitudes.  This may be related to the broad features of H$\alpha$ excess found near the Gum nebula and Orion.
 
\section{Discussion}
\label{sec:conclusion}

The dust content of the Galaxy provides a common origin for both the diffuse FUV and 100 $\mu$m emission.  Cold dust emits at IR wavelengths and efficiently scatters FUV starlight.  In general, these two quantities vary proportionately.  Here we consider two different simplified models of the observed FUV intensity.  The first assumes that it can be modeled as a linear function of 100 $\mu$m emission.  A second refined model fits FUV as a function of Galactic latitude, albedo, and scattering asymmetry, using a modified cosecant fit to overcome the deficiencies described in Section \ref{sec:galactictrends}.  

\subsection{Linear fit between FUV and 100 $\mu$m}

Linearity between FUV intensity and 100 $\mu$m emission is found for points with 100 $\mu$m emission less than $\sim$8 MJy/sr.  Line fits and other evidence discussed in Section \ref{sec:fuvvsall} also indicate a FUV offset at the 100 $\mu$m zero-point.  These two values, the slope and offset, of the linear fit between FUV intensity and 100 $\mu$m emission are discussed in further detail.

\subsubsection{FUV offset}

An FUV offset in the linear relationship with other tracers of the cold Galactic ISM has been observed previously, and in our work appears to be $\sim$300 CU.  This offset is assumed to result from a local source of diffuse isotropic background, an extragalactic background or an isotropic Galactic source not yet considered.  As discussed above in Section 2, some contribution may also come from incomplete masking of known resolved objects.

As the FUV background shows low-level variability over the course of an orbital night, some contribution is likely to originate from \textsc{OI} (1356\AA, 1304\AA) airglow and/or geocoronal Ly $\alpha$ (1216\AA)  lines, which have night-sky intensities of 1.0, 10 and 3000 R respectively.  \textsc{OI} 1356\AA\ falls within the FUV bandpass (at 35\% peak efficiency) resulting in a count rate at the detector corresponding to a $\sim$150 CU FUV continuum background.   GALEX included a blue edge filter which is expected to attenuate the contribution from \textsc{OI} 1304\AA\ and Ly$\alpha$ below these levels. During orbital night we observe a variation in the background intensity of $\pm$50 CU.  Observations of an identical target throughout the year show a similar 50 CU scatter, presumably due to seasonal variation in orbit geometry and airglow intensity.  \citet{2013Murthy} also calculated the expected contribution to the GALEX FUV channel from airglow as a function of both time from local midnight and angle between the Sun and the observed target.  From this work, airglow was estimated to be 200 CU $\pm$ 100 CU at local midnight, comparable to our assessment, with a similar variation vs. local midnight.  As the low level of variation is not likely to impact our analysis, we have left more detailed modelling (and subtraction) of the variable airglow component to a subsequent paper.  It is worth nothing that \citet{2011Seona} found a similar offset using FIMS/SPEAR data while excluding the \textsc{OI} airglow line.  The contribution due to Zodiacal light is sufficiently low for the FUV band that we do not attempt to remove it \citep{1998Leinert}.  Inspection of FUV intensity vs. ecliptic latitude shows no evidence for zodiacal contamination.

Another possible source of low level intensity is unresolved or incompletely masked FUV objects.  There are at least three potential contributors here: 1) unmasked scattered light or ghosts from bright stars; 2) any unresolved and/or undetected light from faint stars or extragalactic objects that have not been masked and 3) unmasked light from the other masked objects in the field.  The GALEX pipeline mask, which is used to remove bright objects (see Section \ref{subsec:mosaic}) could potentially have missed faint stars.  Furthermore unmasked reflections and ghosts around bright stars are visible in the data.  These can contribute to the overall scatter, aside from any contribution to the offset.   In general GALEX avoided observing bright stars with m$_{FUV}\sim$9.5 with a flux producing a local count rate exceeding 5,000 counts per second.  If we conservatively assume that 1\% of the light from the brightest observable star filled an 11 square arcminute pixel then we would anticipate a diffuse contribution of $<$6000 CU at that location on the sky.  The source density of stars just below the avoidance limit (e.g., with 9.5$<m_{FUV}<$12) is low, much less than 1 per square degree over the AIS region, suggesting that fewer than 0.1\% of all pixels may be contaminated by unmasked bright starlight.  Additionally, our object detection software treats most bright stars as extended sources, creating a larger masked area than for fainter unresolved objects.  

Extragalactic diffuse FUV intensity is believed to contribute only a few tens to 100 CU.  \citet{2005Xu} calculated the contribution to the GALEX data from both resolved and unresolved galaxies to be 1.03 $\pm$0.15 nW m$^{-2}$ sr$^{-1}$, or about 51.5 $\pm$7.5 CU.  \citet{2011Voyer} find that the integrated light from field galaxies contribute flux at the level of 65-82 CU to the extragalactic background.   \citet{2011Seonb}, also calculated that the cumulative effect of unmasked unresolved FUV stars and galaxies is probably not significant.  These same results suggest that unmasked light from the objects below the AIS detection limit will be negligible.

Additional components could come from other sources of FUV intensity, including molecular hydrogen fluorescence and line emission such as C IV.  It seems unlikely that there is enough evenly distributed molecular hydrogen at these high latitudes to contribute significantly to the continuum offset.  \citet{2006Ryu} and  \citet{2008Ryu} report band-averaged I(H$_2)/I_{cont}$ ratios of $\sim$0.15 in molecular-rich star-forming; the ratios in diffuse gas is likely to be lower (e.g. \citealt{1990Martin,2006Lee,2008Lee}).   \citet{2008Ryu} also suggest that the band-averaged contribution from C IV is even lower. As significant concentrations of molecular gas are present closer to the disk, H$_2$ fluorescence may contribute to the large scatter for $|$b$|<$25$^\circ$.  

A last concern is whether a possible systematic zero-point offset exists in the comparison data sets.  We can investigate this possibility by comparing the relationship between different tracers of the Galactic ISM, provided that they are uncorrelated.   More recent data sets, such as the Planck map of cold Galactic dust, could provide new measures of the lowest dust column densities.  However a preliminary inspection of the 2013 Planck data indicates the low dust regions are still present at the levels observed previously \citep{2013Planck}.  The offset calculated using these data did not change.

H$\alpha$ provides a different view of the ISM which is indeed suggested in the correlations we observe.  The presence of HII regions will introduce additional scatter, and the intercepts of the fits in these regions are typically a few hundred CU above the value used here.  However, at high latitudes where star-forming regions are not present, the offset decreases to $\sim$400 CU, similar to the offset obtained using other Galactic quantities.  

\subsubsection{FUV-IR Slope}

With the offset removed, the slope of the FUV vs. 100 $\mu$m relation is variable across the sky, as seen in Figures \ref{fig:allskyfuvir} and \ref{fig:slope_lat}.  There are two regimes in the behavior of the FUV.   In the optically thin regime, typically where 100 $\mu$m is less than 8 MJy/sr, the FUV and 100 $\mu$m are correlated.  In the optically thick regime, FUV saturates and the correlation disappears.  In the discussion below we only refer to the optically thin regime.  At mid and high latitudes there are very few regions that deviate from a linear relationship, due to an absence of optically thick dust.

A very simple model, assuming isotropic scattering, an average cosecant dust column relation and a constant scale height, predicts a uniform relation between 100 $\mu$m and FUV across the sky.  Instead, the slope declines with increasing latitudes.  This change in slope for optically thin clouds between mid and high latitudes indicates that a simple scattering picture may not be valid.  The emission at 100 $\mu$m decreases at high latitudes, following the csc$|$b$|$ relation, with the simple model suggesting that the FUV intensity should decrease proportionally.  Our results, as in \citet{2011Seonb}, show that the FUV intensity is decreasing faster than expected, leading to a smaller typical value for the slope at high latitudes.

\subsection{Modified cosecant fit and scattering properties}

The changing slope between FUV intensity and 100 $\mu$m emission at high latitudes is related to the deviations from a cosecant dependence (Figure \ref{fig:fuvsin}).  As discussed in Section \ref{sec:galactictrends}, a function of the form I=A/sin$|$b$|$ for FUV intensity with the offset removed is not able to fully describe the observed intensity.  Adding an extra term to the function, making it I=A/sin$|$b$|$+D, yields better fits for $|$b$| >$ 25$^\circ$ (see also \citealt{2011Seonb}).  But under the assumption that all isotropic components have been accounted for and removed, there is no physical basis for the inclusion of the constant D.

The simple cosecant fit does not include parameters for non-isotropic dust scattering, instead assuming that the dust scattering scale factor was constant with latitude.  \citet{1979Jura} proposed that the surface brightness of a cloud at various latitudes is a function not just of the ISRF, but also of the scattering function for the dust, assuming all illumination originates in the plane.  With the inclusion of optical depth by \citet{1992Wright}, the dust scattered intensity, S, can be expressed as:

\begin{equation}
S = S_o \tau a (1 - 1.1 g \sqrt{sin|b|})
\label{eq:scatter}
\end{equation}

This approximation is valid for $|$b$| >$ 10$^\circ$, g $<$ .85, and optical depths of less than 1 where S$_o$ is the peak scattered ISRF.  To calculate optical depth, we used E(B-V) values from \citet{1998Schlegel}, and a standard optical depth calculation:

\begin{equation}
\tau = \frac{R_{\lambda}}{1.086} \times E(B-V)
\label{eq:scatter2}
\end{equation}

\begin{figure}
\centering
\includegraphics[width=0.45\textwidth]{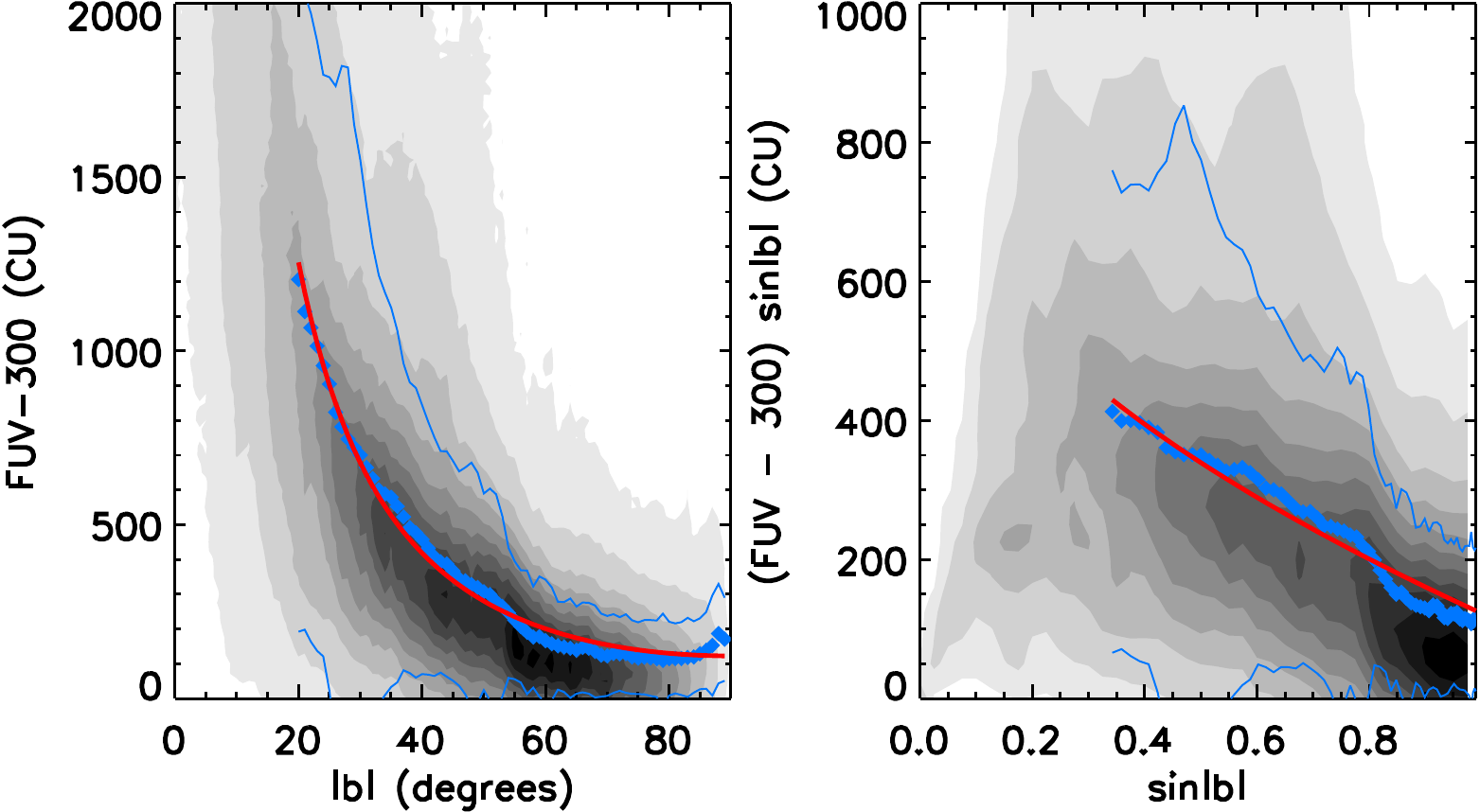}
\caption{2-D histograms of FUV intensity vs. Galactic latitude, with 300 CU offset removed. \textbf{Left Plot:} FUV intensity vs. Galactic latitude, $|$b$|$.  \textbf{Right Plot:} FUV sin$|$b$|$ vs. sin$|$b$|$.  Blue dots are the median for bins of 1$^\circ$, above $|$b$|$=20$^\circ$, with blue lines indicating the standard deviation.  The red line is the best fit of Equation \ref{eq:scatter} for FUV intensity. Values of 0.62 $\pm$ 0.04, 0.78 $\pm$ 0.05, and 6260 $\pm$ 400 are used, for $a$, $g$, and $S_o$, respectively.}\label{fig:scatterplot}
\end{figure}

We then use Equation \ref{eq:scatter} and fit for values of $a$, $g$, and S$_o$.  We compare predicted intensity to FUV intensity (with 300 CU offset removed), and find the best fit values are 0.62 $\pm$ 0.04 and 0.78 $\pm$ 0.05, for $a$ and $g$ respectively, with S$_o$ =6260 $\pm$ 400 CU for sin$|$b$| >$ 0.3, or $|$b$| >$ 20$^\circ$.  Here, a non-linear least squares fit was applied to the median of FUV intensity in bins of $|$b$|$ = 1.0 degree.  There is some degeneracy between the choice of peak intensity S$_o$ and albedo $a$.  The albedo value we predict is related in part to the selection of S$_o$.  Larger values of S$_o$ allow for a smaller $a$.  If we force S$_o$ = 5000, the best fit model predicts an albedo of .75 $\pm$ 0.04, while $g$ remains unchanged.  The value of S$_o$ in the best fit model is similar to the 5800 CU scaling used by \citet{1991Hurwitz}.  An overlay of the fit on the data is shown in Figure \ref{fig:scatterplot}.

As with the cosecant fit, this modified fit is not valid at low latitudes.  The best fit we find is similar to the two part cosecant fit described in Section \ref{sec:galactictrends}, but is able to more accurately capture the effect of asymmetrical scattering.

The values we derive for $a$ and $g$ are slightly higher but within the limits of previous measurements.  \citet{2003Draine}, in a review of previous work, reported a range of modeled values for diffuse Galactic light (DGL) plus values predicted from dust models: $a$ varied between 0.2 and 0.6 in the FUV, while $g$ varied between 0.0 and 0.8.  Albedo from the dust models of \citet{2001Weingartner} was $a$ $\approx$ 0.4, with predicted scattering asymmetry of $g$ $\approx$ 0.7.  \citet{1994Witt} found $a =$ 0.5 and $g =$ 0.9.  \citet{2001Schiminovich} fit for values of $a =$ 0.45 $\pm$ 0.05 and $g =$ 0.77 $\pm$ .1.  \citet{2011Murthy} found limits on $g$ between 0.58 $\pm$ .12, based on scattering angles around individual stars.  \citet{2008Lee} find an albedo of $a =$ 0.36 $\pm$ 0.20, and $g =$ 0.52 $\pm$ 0.22.  Reported values for $a$ and $g$ for individual regions or clouds have lower values than those of the DGL.  For example, \citet{2013Lim} found $a =$ 0.42 $\pm$ 0.05 and $g =$ 0.47 $^{+.11} _{-.27}$, for Taurus-Auriga-Perseus complex.  See \citet{2003Draine} for additional clouds and reflection nebulae.  The range of values for $a$ and $g$ in individual regions reflect in part the ambiguity of geometry in these clouds.  Values for $g$ in particular can vary if the dust is placed behind the illuminating stars or between the stars and the observer.  

\subsection{Deviations in FUV intensity}

After accounting for the FUV offset and the scattering asymmetry of Galactic dust grains, there remain two high latitude regions of the sky that are worth further consideration.  These two regions, one in the northern hemisphere and one in the southern hemisphere, have high FUV/IR ratios ($\sim$600 CU/(MJy/sr)) compared to ratios found in areas at the same latitude ($\sim$200 CU/(MJy/sr)).

As shown in Figure \ref{fig:allskyfuvir}, the region above b=30$^\circ$, and between 60$^\circ$ $< l <$ 120$^\circ$ in the northern hemisphere and below b=-30$^\circ$ and between 240$^\circ$ $< l <$ 300$^\circ$ in the southern hemisphere, have somewhat elevated FUV intensity given the dust content, even with the offset removal.  These regions do partially coincide with structures in the halo (high latitude Galactic clouds) and some Galactic satellites including the Magellanic clouds and stream, but it is unlikely that excess FUV from these objects is being detected.  Uniformly increasing the offset to require that these regions of the sky have zero FUV intensity (as the low levels of N$_{\rm HI}$ and 100 $\mu$m emission might require) ultimately causes other higher latitude regions of the sky to have negative FUV intensities.  Thus, we do not think that the excess FUV intensity in these regions is a result of underestimating the offset.

Instead, we can consider two explanations for the excess FUV intensity here.  The first case is that these regions contain additional FUV intensity that is unrelated to the dust content.  There are well known super bubble regions which contain FUV line emission due to ionized gas and other sources.  However, known super bubbles are not coincident with the regions of interest, instead being adjacent (Orion-Eridanus in the south and Ophiuchus in the north) and likely unrelated to the FUV emission found in these regions.  The known line emission in super bubble regions is estimated at around 32700 LU (in the case of Orion-Eridanus, \citealt{2011Jo,2006Kregenow}), which would translate to 87 CU in the GALEX FUV band.  It can't be ruled out that some additional line emission is contributing to the slight FUV excess in these regions, potentially excited by scattering or remnant shocks from the super bubble.

It might be more natural to consider a second possibility that infers a causal link between the low dust content and enhanced FUV intensity.  One explanation for this may be that these regions are the remnants of old superbubbles.  Such a remnant would have a low dust content as the super-bubble cleared out the dusty ISM in the region although a significant population of older UV-bright stars might remain to illuminate the surrounding area. That this could be the case is loosely suggested by the distribution of OB and A type stars from TD-1, as shown in Figure 15 of \citet{2011Seonb} and the lower ratios of H$\alpha$/FUV which may indicate the presence of a softer local radiation field with fewer ionizing photons.  Furthermore, the presence of holes or chimneys in these directions is also suggested by observations of the 3-D distribution of the Local Bubble \citep{2003Lallement} and the link to the EUV/soft-X-ray background.  Nearby nebular regions in the disk (e.g. the Gum Nebula) remain poorly understood and have been shown to contain expanding gas and clouds \citep{2001Woermann} that may also trace successive generations of star formation.  We note that while star formation rate (e.g. FUV starlight) and molecular gas are globally correlated in external galaxy disks, on smaller physical scales less than 1 kpc they are known to show considerable scatter (e.g. \citet{2010Schruba}, \citet{2013Leroy}).

\section{Summary}
\label{sec:summary}

We have constructed an all sky map of diffuse Galactic FUV intensity, using GALEX FUV AIS data, covering 65\% of the sky.  We have compared our map to other maps of the diffuse FUV sky and to maps of complementary Galactic quantities.  We find the FUV intensity is highly dependent on a combination of 100 $\mu$m emission, Galactic latitude, and proximity to UV bright stars and OB associations.  

Our main conclusions are:

\begin{enumerate}
\item FUV intensity is highest near the Galactic plane and around known OB associations and lowest at high latitudes.
\item There is a $\sim$300 CU FUV isotropic offset which is likely due to a combination of air glow (likely the dominant contributor), a small extragalactic background component including continuum light from unresolved galaxies, and/or a Galactic component not traced by other indicators.
\item FUV intensity and 100 $\mu$m emission show a  linear correlation below 8 MJy/sr of 100 $\mu$m.
\item FUV intensity and N$_{\rm HI}$ show a linear correlation below 1.2 $\times$ 10$^{21}$ cm$^{-2}$.
\item FUV intensity follows a modified cosecant shape with Galactic latitude with low intensity at high latitudes due to strongly forward scattering dust grains.
\item We calculate a best fit value of $g$=0.78 $\pm$0.05 for the scattering asymmetry, with $a$=0.62 $\pm$0.04 for albedo, and a peak scattering intensity, S$_o$=6260 $\pm$ 400 CU, for all points with $|$b$|> $20$^\circ$.
\end{enumerate}

A simple picture of this behavior is that the direct, linear variation of FUV intensity at low 100 $\mu$m emission can be explained as scattered starlight off of low optical depth dust.  As 100 $\mu$m emission increases, the FUV intensity increases until reaching a plateau where the dust begins to self-shield.  This plateau occurs at around 8 MJy/sr and appears to be constant across the sky.  The exact ratio of FUV to 100 $\mu$m emission appears to depend on Galactic latitude, with starlight more effectively scattered at lower latitudes.  The scatter in diffuse FUV intensity across a single latitude is primarily caused by anisotropies in the interstellar radiation field, including the scale heights of different stellar components, and geometrical effects caused by the exact structure of individual dust clouds.  Less important, but still a component of the scatter could be variations in the type of dust present and the properties of that dust.

Of further interest are individual, small regions (less than 1 degree across) where FUV intensity deviates from the expected linear relationship with 100 $\mu$m intensity and modified cosecant model with Galactic latitude.  These regions are typically individual dusty clouds or groups of clouds with high 100 $\mu$m emission.  While only 10 \% of our data covers points with 100 $\mu$m emission above 8 MJy/sr, this still contains numerous regions with flat or even inverse relationships between FUV and 100 $\mu$m.  In a future paper, we discuss in detail individual clouds that deviate from the models described above, some of which show evidence for FUV obscuration or excess FUV intensity.  

\acknowledgements
The authors wish to thank the anonymous reviewer for their detailed and constructive comments. This publication is based on observations made with the NASA Galaxy Evolution Explorer. GALEX was operated for NASA by the California Institute of Technology under NASA contract NAS5-98034.  We acknowledge the use of the Legacy Archive for Microwave Background Data Analysis (LAMBDA), part of the High Energy Astrophysics Science Archive Center (HEASARC). HEASARC/LAMBDA is a service of the Astrophysics Science Division at the NASA Goddard Space Flight Center.  This research made use of Montage, funded by the National Aeronautics and Space Administration's Earth Science Technology Office, Computation Technologies Project, under Cooperative Agreement Number NCC5-626 between NASA and the California Institute of Technology. Montage is maintained by the NASA/IPAC Infrared Science Archive.  The authors also wish to thank Josh Peek for helpful discussions.


\end{document}